\documentclass{elsart}

\usepackage{amsmath}
\usepackage{amssymb}
\usepackage[a4paper]{geometry}
\usepackage{graphicx}
\usepackage{epsfig}

\def\bsigma{\mbox{\boldmath $\sigma$}}

\begin{document}
\topmargin -1cm

\begin{frontmatter}

\title{Parallel dynamics of the fully connected Blume-Emery-Griffiths neural network}
\author{D. Boll\'e \thanksref{Bolle}}
\author{, J. Busquets Blanco \thanksref{Jordi}}
\address{Instituut voor Theoretische Fysica, Katholieke Universiteit Leuven, Celestijnenlaan 200 D, B-3001 Leuven, Belgium}
\author{G.M.Shim \thanksref{Shim}} 
\address{Department of Physics, Chungnam National 
            University ,Yuseong, Daejeon 305-764, R.O.~Korea}
\thanks[Bolle]{desire.bolle@fys.kuleuven.ac.be}
\thanks[Jordi]{jordi.busquets@fys.kuleuven.ac.be}
\thanks[Shim]{gmshim@cnu.ac.kr}
\begin{abstract}
The parallel dynamics of the fully connected Blume-Emery-Griffiths neural network model is studied at zero temperature
using a probabilistic approach.
A recursive scheme is found determining the complete time evolution
of the order parameters, taking into account {\em all} feedback correlations. It is based upon the evolution of the distribution of the local field, the structure of which is determined in detail.
As an illustrative example explicit analytic formula are given for the
first few time steps of 
the dynamics. 
Furthermore, equilibrium fixed-point equations are derived and compared with the thermodynamic approach. The analytic results find excellent confirmation in extensive numerical simulations.

\vspace*{0.5cm}
\end{abstract}
\end{frontmatter}

\section{Introduction}
\label{sec:intro}
Recently, an optimal Hamiltonian has been derived in the statistical mechanics approach to $Q$-state neural networks starting from the concept of mutual information \cite{DK00},\cite{BV02}. Optimal means that the best retrieval properties are guaranteed including, e.g., the  largest retrieval overlap, loading capacity, basin of attraction, convergence time. For $Q=3$ this Hamiltonian resembles the classical Blume-Emery-Griffiths (BEG) Hamiltonian in the sense that it contains both a  bilinear and biquadratic term in the spins \cite{BEG}. For a fully connected architecture it has been shown, using a thermodynamic replica approach that the maximal loading capacity for the BEG network is indeed bigger than the one for other three-state networks existing in the literature (Ising, Potts...) \cite{BV02}. 

The dynamics of these new type of neural network models is under investigation. In the case of an asymmetric extremely diluted architecture where one knows that there are no feedback loops complicating the time evolution, this dynamics has been solved in closed form \cite{DK00} showing a better retrieval quality than other diluted models for certain parameters of the system. 
For symmetric architectures -- fully connected but also extremely diluted -- however, the situation is much more complicated. From former experience with, e.g., $Q$-Ising models (see \cite{BJS99} and references therein) one knows that the dynamics is very non-trivial due to the feedback in the system \cite{BKS90}. This feedback causes the appearance of discrete noise, besides Gaussian noise, involving the neurons at all previous time steps and prevents a closed-form solution. 

In this work we generalize the probabilistic approach that has been developed for the Hopfield model \cite{PZ91}, \cite{PZ91b} and $Q$-Ising model \cite{BVZ93} in order to solve the dynamics for the fully connected BEG model at zero temperature. Thereby, we start from the time evolution of the distribution of the local field, instead of working directly with the order parameters. We study the structure of this distribution in detail and, using this knowledge, we develop a recursive scheme in order to calculate the relevant order parameters of the system, i.e., the main overlap, the neural activity, the activity overlap and the variance of the residual overlap at any time step. 
As an illustration we write out these expressions in detail for the first few time steps of the dynamics. 

Furthermore, by requiring the local field to be time-independent, implying that some correlations between its Gaussian and discrete noise parts are neglected, we derive fixed-point equations for the order parameters. They coincide with those derived via thermodynamical methods \cite{BV02}. 

Finally we perform numerical simulations of a BEG network with $N=6000$ neurons. They confirm the analytical results we have derived. 

The rest of this paper is organized as follows. 
In Section \ref{sec:mod} we introduce the model, its dynamics and the relevant order parameters. In Section \ref{sec:gensch} we use the
probabilistic approach in order to derive a recursive scheme for the
evolution of the distribution of the local field, leading to recursion
relations for the order parameters. 
Using this scheme, we explicitly calculate in Appendix~A the
order parameters and in Appendix~B the local field for the first few time steps of
the dynamics. In Section \ref{sec:fixp} we show the existence of a Lyapunov function at zero temperature and we discuss the evolution of
the system to fixed-point attractors. Section \ref{sec:fields} details the structure of the local field distribution, especially the appearance of gaps. The analytic results are compared with numerical simulations in Section \ref{sec:results}. Some concluding  remarks are given in Section \ref{sec:con}.

\section{The model}
\label{sec:mod}
Consider a neural network consisting of $N$ neurons which can take
values $\sigma_i, i=1,\ldots, N$ from the discrete set
$ \mathcal{S}\equiv \lbrace -1,0,+1 \rbrace $.
The $p=\alpha N$ patterns to be stored in this network are supposed to
be a collection of independent and identically distributed random
variables (i.i.d.r.v.), $\{\xi_i^\mu\} $,
$\mu =1,\ldots,p$ 
with a probability distribution 
\begin{equation}
p(\xi_{i}^{\mu})=\frac{a}{2}\delta(\xi_{i}^{\mu}-1)+\frac{a}{2}\delta(\xi_{i}^{\mu}+1)+(1-a)\delta(\xi_{i}^{\mu})
\end{equation}
with $a$ the activity of the patterns so that
\begin{equation}
  \lim_{N\rightarrow\infty}\frac{1}{N}\sum_{i}(\xi_{i}^{\mu})^2 = a.
\end{equation}

Given the network configuration at time $t$,
${\bsigma}_N(t)\equiv\{\sigma_j(t)\}, j=1,\ldots,N$,
the following dynamics is considered.   
The configuration $\bsigma_N(0)$ is chosen as input. 
At zero temperature all neurons are updated in parallel according to the
rule
\begin{equation}
        \label{eq:enpot}
     \sigma_i(t)\rightarrow\sigma_i(t+1)=s':
         \min_{s\in \mathcal {S}} \epsilon_i[s|{\bsigma}_N(t)]                      =\epsilon_i[s'|{\bsigma}_N(t)]
\end{equation}
with $s' \in \mathcal{S}$.
We remark that this rule is the zero temperature limit of the stochastic parallel spin-flip dynamics defined by the transition probabilities
\begin{equation}
   \Pr \left(\sigma_i(t+1) = s' \in \mathcal{S}| \bsigma_N(t) \right)
        =
        \frac
        {\exp [- \beta \epsilon_i(s'|\bsigma_N(t))]}
        {\sum_{s \in \mathcal{S}} \exp [- \beta \epsilon_i
                                   (s|\bsigma_N(t))]}\,.
\label{eq:trans}
\end{equation}
Here the energy potential $\epsilon_i[s|{\bsigma}_N(t)]$
is defined by 
\begin{equation}
          \epsilon_i(s|{\bsigma}_N(t)) =                                                 -sh_i({\bsigma}_N(t))-s^2\theta_i({\bsigma}_N(t))    
                                            \,,
\end{equation}
where the following local fields in neuron $i$ carry all the information
\begin{equation}
        \label{eq:h}
      h_{N,i}(t)=\sum_{j \neq i} J_{ij}\sigma_j(t), \quad
      \theta_{N,i}(t)=\sum_{j\neq i}K_{ij}\sigma_{j}^{2}(t)
\end{equation}
with the obvious shorthand notation for the local fields. The synaptic couplings $J_{ij}$ and $K_{ij}$ are of the Hebb-type
\begin{equation}
J_{ij}=\frac{1}{a^{2}N}\sum_{\mu=1}^{p}\xi_{i}^{\mu}\xi_{j}^{\mu},\qquad K_{ij}=\frac{1}{N}\sum_{\mu=1}^{p}\eta_{i}^{\mu}\eta_{j}^{\mu}
\end{equation}
with
\begin{equation}
\eta_{i}^{\mu}=\frac{1}{a(1-a)}((\xi_{i}^{\mu})^{2}-a).
\end{equation} 
These are the local fields entering in the BEG model \cite{DK00}. The updating rule (\ref{eq:enpot}) is equivalent to using a gain function 
\begin{equation}
        \label{eq:gain}
        \sigma_i(t+1)  = 
               \mbox{g}(h_{N,i}(t), \theta_{N,i}(t))=
         \mbox{sign}(h_{N,i}(t)) \Theta(|h_{N,i}(t)| + \theta_{N,i}(t))
\end{equation}
with $\Theta$ the Heaviside function.

The order parameters of this system have been obtained starting form the mutual information as a measure for the retrieval quality of the system \cite{DK00}, \cite{BV02}. They are the retrieval overlap, the activity overlap, and  the neural activity 
\begin{equation}
        \label{eq:mdef}
        m_N^\mu(t)=\frac{1}{aN}
                \sum_{i}\xi_i^\mu\sigma_i(t),
                \quad  
        n_N^\mu(t)=\frac{1}{aN}\sum_{i}(\xi_i^\mu)^2(\sigma_i(t))^2,                      \quad
       q_N(t)=\frac{1}{N}\sum_{i}(\sigma_{i}(t))^2 \,.
\end{equation}
Instead of using the activity overlap $n_N^\mu(t)$ itself it is more convenient to employ the modified activity overlap
\begin{equation}
l_N^\mu(t)=\frac{1}{1-a}(n_N^\mu(t) - q_N(t))
  =\frac{1}{N}\sum_{i}(\eta_i^\mu)(\sigma_i(t))^2.
  \label{eq:ldef}
\end{equation}

\section{Recursive dynamical scheme}
\label{sec:gensch}
In networks with symmetric couplings it is known that non-trivial correlations occur, even at zero temperature, which become increasingly tedious to evaluate \cite{BKS90}.

On the basis of the probabilistic approach (see, e.g., \cite{PZ91}, \cite{PZ91b}, \cite{BVZ93}) used successfully before (\cite{BJS99} and references therein) we develop in this section a recursive dynamical scheme in order to study the time evolution of the distribution of the local fields $h_i(t)$ and $\theta_i(t)$. This allows us to write down recursion relations determining the full time evolution of the order parameters (\ref{eq:mdef})-(\ref{eq:ldef}) of the BEG network model.

Suppose that the initial configuration of the network $\{\sigma_i(0)\}$ is a collection of i.i.d.r.v. with mean $\mbox{E}[\sigma_{i}(0)]=0$ and variance $\mbox{Var}[\sigma_{i}(0)]= q_{0} $
and correlated with only one pattern which we choose, without loss of generality, to be the first one 
\begin{equation}
 \label{ini2}
\mbox{E}[\xi_{i}^{\mu}\sigma_{j}(0)]=
 \delta_{i,j}\delta_{\mu,1}m_{0}^{1}a, \quad m_{0}^{1}>0, \quad
\mbox{E}[ \eta_j^{\mu}\sigma_{i}^{2}(0)]=       \delta_{i,j}\delta_{\mu,1}l_0^1 \,.
\end{equation}

By the law of large numbers (LLN) eqs. (\ref{eq:mdef})-(\ref{eq:ldef}) and (\ref{ini2}) determine  the order parameters $m^1_N(0),q_N(0)$ and $l^1_N(0)$ at $t=0$ in the limit $N \rightarrow \infty$.

Next, we want to apply standard signal-to-noise techniques (see, e.g, \cite{PZ91}, \cite{BVZ93}) to both the local fields $h_{N,i}(0)$ and $\theta_{N,i}(0)$ at $t=0$. Starting from their definitions we find 
\begin{eqnarray}
h_{i}(0)
&=&\lim_{N\rightarrow\infty}\Big(\frac{1}{a}\xi_{i}^{1}m_{N}^{1}(0)-
\frac{1}{a^{2}N}(\xi_{i}^{1})^{2}\sigma_{i}(0)+
\frac{1}{a^{2}N}\sum_{\mu>1}\sum_{j\neq i}
\xi_{i}^{\mu}\xi_{j}^{\mu}\sigma_{j}(0)\Big)
            \nonumber \\
            \label{hgauss}
&\stackrel{\mathcal D}{=}&
\frac{1}{a}\xi_{i}^{1}m^1(0)+\mathcal{N}\big(0,\frac{\alpha q(0)}{a^{2}}\big)
            \\      
\theta_{i}(0)&=&
\lim_{N\rightarrow\infty}\Big(\eta_{i}^{1}l_{N}^1(0)-
\frac{1}{N}(\eta_{i}^{1})^{2}\sigma_{i}^{2}(0)+
\frac{1}{N}\sum_{\mu>1}\sum_{j\neq i}
\eta_{i}^{\mu} \eta_{j}^{\mu} \sigma_{j}^2(0) \Big)
             \nonumber \\
             \label{thetagauss}
&\stackrel{\mathcal D}{=}&
\eta_{i}^{1}l^1(0)+\mathcal{N}\big(0,\frac{\alpha q(0)}{a^{2}(1-a)^{2}}\big)\,,         
\end{eqnarray}
where the convergence is in distribution \cite{S}. The quantity $\mathcal{N}(0,d) $ represents a Gaussian random variable with mean $0$ and variance $d$.

The key question is then how these quantities evolve in time under the parallel dynamics specified before. For a general time step we find from eq. (\ref{eq:gain}) and the LLN in the limit $N \rightarrow \infty$ for the order parameters (eqs.(\ref{eq:mdef})-(\ref{eq:ldef}))
\begin{eqnarray}
 \label{defm}
m^{1}(t+1)&\stackrel{\mathcal Pr}{=}&
\frac{1}{a}\left\langle\!\left\langle\xi_i^{1}
    g\big(h_{i}(t),\theta_{i}(t)\big)\right\rangle\!\right\rangle
             \\ \label{defq}
q(t+1)&\stackrel{\mathcal Pr}{=}&
\left\langle\!\left\langle\xi_i^{1}g^{2}\big(h_{i}(t),\theta_{i}(t)
       \big)\right\rangle\!\right\rangle
\\ \label{defl}
l^{1}(t+1)&\stackrel{\mathcal Pr}{=}&
\left\langle\!\left\langle\eta_i^1 g^{2}\big(h_{i}(t),\theta_{i}(t)\big)\right\rangle\!\right\rangle ,
\end{eqnarray}
where $h_{i}(t)=\lim_{N\rightarrow\infty}h_{N,i}(t)$  (with an analogous formula for $\theta_{i}(t)$), and where the convergence is in probability. In the above $\left\langle \!\left\langle\cdot \right\rangle\!\right\rangle$ denotes the average both over the distribution of the $\{\xi_i^{\mu}\}$ (and hence $\{\eta_i^{\mu}\}$) and the $\{\sigma_i(0)\}$. Note that the average over the latter is hidden in an average over the local field through the updating rule (\ref{eq:gain}). From the work on symmetric $Q$-Ising networks \cite{BJS99}, \cite{BJS98} we know that due to the correlations we have to study carefully the influence of non-condensed ($\mu >1$) patterns in the time evolution of the system, expressed by the variance of the residual overlaps, in our case in both the local fields. The latter are defined as  
\begin{eqnarray}
r^{\mu}(t)&\equiv&\lim_{N\rightarrow\infty}r_{N}^{\mu}(t)
     =\lim_{N\rightarrow\infty}\frac{1}{a^2\sqrt{N}}
                       \sum_{j}\xi_{j}^{\mu}\sigma_{j}(t),
             \quad \mu > 1   \\
s^{\mu}(t)&\equiv&\lim_{N\rightarrow\infty}s_{N}^{\mu}(t)
      =\lim_{N\rightarrow\infty}\frac{1}{\sqrt{N}}
                       \sum_{j}\eta_{j}^{\mu}\sigma^2_{j}(t),
             \quad \mu > 1
\end{eqnarray}
where the limit $N \rightarrow \infty$ of $r_N^{\mu}(0)$ and $s_N^{\mu}(0)$ is given by the Gaussian random variable in eqs. (\ref{hgauss}) and (\ref{thetagauss}).
At this point we want to  remark that the choice of the initial configurations assures the independence of  $r^{\mu}(0)$ and $s^{\mu}(0)$ as can be seen by calculating the characteristic function $\mbox{E}[\exp(ixr_{N}^{\mu}(0)+iys_{N}^{\mu}(0))]$. 

The further aim of this section is then to calculate the distribution of the local fields and the order parameters as a function of time.

We start by rewriting the local fields (\ref{eq:h}) at time $t$ in the following way
\begin{eqnarray}
h_{N,i}(t)&=&\frac{1}{a}\xi_{i}^{1}m_{N}^{1}(t)+\frac{1}{a^{2}N}\sum_{\mu>1} \sum_j
\xi_{i}^{\mu}\xi_{j}^{\mu}\sigma_{j}(t)-\frac{\alpha}{a}\sigma_{i}(t)
          \nonumber  \\
   &=&\frac{1}{a}\xi_{i}^{1}m_{N}^1(t)-\frac{\alpha}{a}\sigma_{i}(t)+
    \frac{1}{\sqrt{N}}\sum_{\mu>1}\xi_{i}^{\mu}r_{N}^{\mu}(t)
          \label{lfr}  \\
\theta_{N,i}(t)&=&\eta_{i}^{1}l_{N}^1(t)+
\frac{1}{N}\sum_{\mu>1}\sum_{j}
\eta_{i}^{\mu} \eta_{j}^{\mu} \sigma_{j}^2(t)-\frac{\alpha}{a(1-a)}\sigma_{i}^{2}(t)
           \nonumber  \\
   &=& \eta_{i}^{1}l_{N}^1(t)-\frac{\alpha}{a(1-a)}\sigma_{i}^{2}(t)+
    \frac{1}{\sqrt{N}}\sum_{\mu>1}\eta_{i}^{\mu}s_{N}^{\mu}(t)\,.
    \label{lfs}
\end{eqnarray}
From a technical point of view the explicit addition and subtraction of the $\sigma_{i}(t)$ ($\sigma_{i}^{2}(t)$) term is convenient in order to treat all indices in the sum over $j$ on equal footing, which is important to take into account all possible feedback loops.

At this point several remarks are in order. 
Since the neuronal states $\{\sigma_j(t)\}$, for $t>0$, are not i.i.d.r.v., the
central limit theorem (CLT) can not be applied directly to the residual
overlap $r_N^\mu(t)$ and $s_N^\mu(t)$. Furthermore, the set of $\alpha N$ variables $\{\xi_{i}^{\mu}r_{N}^{\mu}(t)\}_{\mu}$ and $\{\eta_{i}^{\mu}s_{N}^{\mu}(t)\}_{\mu}$ are not independent because the $r_{N}^{\nu}(t)$ respectively $s_N^\nu(t), \nu \neq \mu$ are weakly dependent on the $\xi_{i}^{\mu}$ respectively $\eta_{i}^{\mu}$. Indeed, after applying the dynamics, the $\sigma_{i}(t)\,$ ($\sigma^2_{i}(t)$) and the $\xi_{i}^{\mu}\,(\eta_{i}^{\mu})$ become dependent, leading to a weak depence of $r_{N}^{\mu}(t)\,(s_{N}^{\mu}(t))$ and $\xi_{i}^{\mu}\,(\eta_{i}^{\mu})$. This microscopic dependence gives rise to a macroscopic contribution after summing and taking the limit $N \rightarrow \infty$. Therefore, we follow a procedure similar to the one used in the $Q$-Ising model \cite{BJS99}, \cite{BJS98} by isolating in the local fields precisely the contributions arising from these dependences.

In order to do so we rewrite the residual overlaps as 
\begin{eqnarray}
        \label{rmod}
r_{N}^{\mu}(t+1)&=&\frac{1}{a^{2}\sqrt{N}}
    \sum_{i}\xi_{i}^{\mu}g\big(\tilde{h}_{N,i}^{\mu}(t)
   +\frac{1}{\sqrt{N}}\xi_{i}^{\mu}r_{N}^{\mu}(t),
                                \tilde{\theta}_{N,i}^{\mu}(t)
   +\frac{1}{\sqrt{N}}\eta_{i}^{\mu}s_{N}^{\mu}(t)\big)
            \\
            \label{smod}
s_{N}^{\mu}(t+1)&=&\frac{1}{\sqrt{N}}
    \sum_{i}\eta_{i}^{\mu}g^{2}\big(\tilde{h}_{N,i}^{\mu}(t)
   +\frac{1}{\sqrt{N}}\xi_{i}^{\mu}r_{N}^{\mu}(t),
                                \tilde{\theta}_{N,i}^{\mu}(t)
+\frac{1}{\sqrt{N}}\eta_{i}^{\mu}s_{N}^{\mu}(t)\big)\,
\end{eqnarray}
with obvious notation. In these expressions we have extracted the contribution of the $\mu$ term out of the local fields such that the modified local fields $\tilde{h}_{N,i}^{\mu}(t)$ and $\tilde{\theta}_{N,i}^{\mu}(t)$ are only weakly dependent on $\xi_{i}^{\mu}$ and $\eta_{i}^{\mu}$ respectively, whereas $h_{N,i}^{\mu}(t)$ and $\theta_{N,i}^{\mu}(t)$ depend strongly on them. 

Next, we want to find the most important terms in (\ref{rmod}) and (\ref{smod}) in the limit $N \rightarrow \infty$. Therefore, we consider  the characteristic function $\mbox{E}[\exp(ixr_{N}^{\mu}(t+1)+iys_{N}^{\mu}(t+1))]$ using (\ref{rmod}) and (\ref{smod}), up to order $\mathcal{O}(N^{-3/2})$. We then expand the gain function around the modified local fields. After some calculation we obtain in the limit $N \rightarrow \infty$
\begin{eqnarray}
&&\lim_{ N \rightarrow \infty}
\mbox{E}[\exp(ixr_{N}^{\mu}(t+1)+iys_{N}^{\mu}(t+1))] \nonumber \\
 &&  = \exp\left[ix\chi_{h}(t)r^{\mu}(t) -x^2 \frac{q(t+1)}{2a^3}
       +iy \chi_{\theta}(t) s^{\mu}(t) -y^2 \frac{q(t+1)}{2a(1-a)}
               \right] \label{charac}
\end{eqnarray}
with $\chi_{h}(t)$ and $\chi_{\theta}(t)$ the ``susceptibilities'' corresponding to the fields $h_i(t)$ and $\theta_i(t)$ and given by
\begin{eqnarray}
\chi_{h}(t)=\left\langle\!\left\langle
         \frac{\partial g}{\partial h}\Big|_{\tilde{h},\tilde{\theta}}
                 \right\rangle\!\right\rangle
         &=&\left\langle\!\left\langle
\frac{1}{a}\int_{-\infty}^{\infty}d\hat{h}\int_{-\infty}^{\infty}
  d\hat{\theta}
  \rho_{\tilde{h}(t)}(\hat{h})\rho_{\tilde{\theta}(t)}(\hat{\theta})
\Big(\frac{\partial g}{\partial h}\Big|_{\hat{h},\hat{\theta}}\Big)
     \right\rangle\!\right\rangle
           \label{chih} \nonumber \\
       &=&\frac{2}{a}\left\langle\!\left\langle                                \int_{0}^{\infty}d\hat\theta\rho_{\tilde{h}(t)}(0)
       \rho_{\tilde{\theta}(t)}(\hat\theta)\right\rangle\!\right\rangle 
              + (1-a) \chi_{\theta}(t)
           \\
\chi_{\theta}(t)=\left\langle\!\left\langle
  \frac{\partial g^{2}}{\partial\theta}\Big|_{\tilde{h},\tilde{\theta}}
     \right\rangle\!\right\rangle
&=&\frac{1}{a(1-a)}\left\langle\!\left\langle                              \int_{-\infty}^{0}d\hat\theta\rho_{\tilde{\theta}(t)}(\hat\theta)
            (\rho_{\tilde{h}(t)}(\hat\theta)+ \rho_{\tilde{h}(t)}(-\hat\theta))
    \right\rangle\!\right\rangle\,.
    \label{chit}
\end{eqnarray}
In these expressions, $\rho_{\tilde{h}(t)}( h)$ and $\rho_{\tilde{\theta}(t)}( \theta)$ are the probability densities of the modified local fields, $\tilde{h}$ and $\tilde{\theta}$. They are the integrations of the joint distribution $\rho_{\tilde{h}(t),\tilde {\theta}(t)}( h,\theta)$ over the $ h$ and $ \theta$ values, e.g. $\rho_{\tilde{h}(t)}( h)= \int d  \theta \rho_{\tilde{h}(t),\tilde {\theta}(t)}( h,  \theta)$. 
(See  Section \ref{sec:fields} for more details).
From the expansion (\ref{charac}) we see that the local fields $h_{N,i}(t)$ and $\theta_{N,i}(t)$ are independent up to the order $\mathcal{O}(N^{-3/2})$ since $r^{\mu}_N(t)$ and $s^{\mu}_N(t)$ are as well. 

Identifying terms we then get
\begin{eqnarray}
r^{\mu}(t+1)&=&
            \tilde{r}^{\mu}(t)+\chi_{h}(t)r^{\mu}(t)
     \label{recr}         \\
s^{\mu}(t+1)&=&
           \tilde{s}^{\mu}(t)+\chi_{\theta}(t)s^{\mu}(t)\,,
           \label{recs}
\end{eqnarray}
where
\begin{eqnarray}
\tilde{r}^{\mu}(t)
&=&\lim_{N\rightarrow\infty}\frac{1}{a^{2}\sqrt{N}}\sum_{i}\xi_{i}^{\mu}
  g\big(\tilde{h}_{N,i}(t),\tilde{\theta}_{N,i}(t)\big)
\stackrel{\mathcal D}{=}\mathcal{N}\big(0,\frac{1}{a^{3}}q(t+1)\big), \\
\tilde{s}^{\mu}(t)
&=&\lim_{N\rightarrow\infty}\frac{1}{\sqrt{N}}\sum_{i}\eta_{i}^{\mu}
     g^2\big(\tilde{h}_{N,i}(t),\tilde{\theta}_{N,i}(t)\big)
\stackrel{\mathcal D}{=}\mathcal{N}\big(0,\frac{1}{a(1-a)}q(t+1)\big)\,. 
\end{eqnarray}
We remark that this calculation also shows us that $r^\mu(t)$ and $s^\mu(t)$ are independent for all times.

In this way we obtain in the limit $N \rightarrow \infty$ from eqs.(\ref{lfr}) and (\ref{lfs})
\begin{eqnarray}
&&h_{i}(t+1)
=\frac{1}{a}\xi_{i}^{1}m^1(t+1)
   +\chi_{h}(t)\Big\{h_{i}(t)-\frac{1}{a}\xi_{i}^{1}m^1(t)
    +\frac{\alpha}{a}\sigma_{i}(t)\Big\} 
    \nonumber \\
    && \hspace*{2cm}
    +\mathcal{N}\Big(0,\frac{\alpha}{a^{2}}q(t+1)\Big)
    \label{rech}  \\
&&\theta_{i}(t+1)
=\eta_{i}^{1}l^1(t+1)
  +\chi_{\theta}(t)\Big\{\theta_{i}(t)-\eta_{i}^{1}l^1(t)
       +\frac{\alpha}{a(1-a)}\sigma_{i}^{2}(t)\Big\} 
       \nonumber \\
      && \hspace*{2cm}
      +\mathcal{N}\Big(0,\frac{\alpha}{a^{2}(1-a)^{2}}q(t+1)\Big)\,.
    \nonumber\\  \label{rect}
\end{eqnarray}

From this it is clear that the local fields at time $t+1$ consist out of a discrete part and a normally distributed part, viz.
\begin{eqnarray}
h_{i}(t+1)&=&M_{i}(t+1)+\mathcal{N}\big(0,V(t+1)\big)\\
\theta_{i}(t+1)&=&L_{i}(t+1)+\mathcal{N}\big(0,W(t+1)\big)\,,
\end{eqnarray}
where
\begin{eqnarray}
M_{i}(t+1)&=&\chi_{h}(t)\Big[M_{i}(t)-\frac{\xi_{i}^{1}}{a}m^1(t)+\frac{\alpha}{a}\sigma_{i}(t)\Big]+\frac{\xi_{i}^{1}}{a}m^1(t+1)
   \label{recM} \\
L_{i}(t+1)&=&\chi_{\theta}(t)\Big[L_{i}(t)-\eta_{i}^{1}l^1(t)+\frac{\alpha}{a(1-a)}\sigma_{i}^{2}(t)\Big]+\eta_{i}^{1}l^1(t+1)
      \label{recL}
\end{eqnarray}
and
\begin{equation}
V(t+1)=\alpha aD(t+1), \quad
W(t+1)=\frac{\alpha}{a(1-a)}E(t+1)
       \label{VWini}
\end{equation}
with $D(t+1)$ and $E(t+1)$ the variances of the residual overlaps, $r^{\mu}(t+1)$ and $s^{\mu}(t+1)$, satisfying the recursion relations 
\begin{eqnarray}
 D(t+1)&=&\frac{q(t+1)}{a^{3}}+\chi_{h}^{2}(t)D(t)
          +2\chi_{h}(t)\mbox{Cov}[\tilde{r}^{\mu}(t), r^{\mu}(t) ]
  \label{recD} \\
 E(t+1)&=&\frac{q(t+1)}{a(1-a)}+\chi_{\theta}^{2}(t)E(t)
            +2\chi_{\theta}(t)\mbox{Cov}[\tilde{s}^{\mu}(t), s^{\mu}(t) ] \,.               \label{recE}
\end{eqnarray}
We still have to determine $\rho_{\tilde{h}(t)}(h)$ and $\rho_{\tilde{\theta}(t)}(\theta)$ in (\ref{chih}) and (\ref{chit}). We know that the quantities $M_{i}(t)$ and $L_{i}(t)$ consist out of a signal term and a discrete noise term, viz.
\begin{eqnarray}
M_{i}(t)&=&\frac{\xi_{i}^{1}}{a}m^1(t)+\sum_{t'=0}^{t-1}\frac{\alpha}{a}\Big[\prod_{s=t'}^{t-1}\chi_{h}(s)\Big]\sigma_{i}(t')\\
L_{i}(t)&=&\eta_{i}^{1}l^1(t)+\sum_{t'=0}^{t-1}\frac{\alpha}{a(1-a)}\Big[\prod_{s=t'}^{t-1}\chi_{\theta}(s)\Big]\sigma_{i}^{2}(t').
\end{eqnarray}

The evolution equation tells that the $\sigma_i(t')$ and $\sigma_i^{2}(t')$ can be written in terms of the $h_{i}(t'-1)$ and $\theta_{i}(t'-1)$ such that the second terms in the expressions above are the sum of correlated variables. Furthermore, these are also correlated through the dynamics with the normally distributed part of the local fields. So the local fields can be considered as a transformation of a set of correlated variables ${\bf x}=\{x_s\}, {\bf y}=\{y_s\}, s=1,2, \ldots, t-2,t$  which we choose to normalise. Then we arrive at the following expression for the probability densities of the local fields 
\begin{eqnarray}
\label{rhoh}
\lim_{N\rightarrow\infty}\rho_{\tilde{h}_{i}^{\mu}(t)}(h)
            =\rho_{h_{i}(t)}(h)
     &=&\int\big(\prod_{s=0}^{t-2}dx_{s}dy_{s}\big)dx_{t}dy_{t}
 \frac{1}{\sqrt{\mbox{det}(2\pi C_{h})}}\frac{1}{\sqrt{\mbox{det}(2\pi C_{\theta})}}
              \nonumber \\
&&\exp\big(-\frac{1}{2}{\bf x}C_{h}^{-1}{\bf x}^{T}
           -\frac{1}{2}{\bf y}C_{\theta}^{-1}{\bf y}^{T}\big)
\delta\big(h-M_{i}(t)-\sqrt{V(t)}x_{t}\big)
               \nonumber \\ \\
\label{rhot}
\lim_{N\rightarrow\infty}\rho_{\tilde{\theta}_{i}^{\mu}(t)}(\theta)
               =\rho_{\theta_{i}(t)}(\theta)
       &=&\int\big(\prod_{s=0}^{t-2}dx_{s}dy_{s}\big)dx_{t}dy_{t}
  \frac{1}{\sqrt{\mbox{det}(2\pi C_{h})}}\frac{1}{\sqrt{\mbox{det}(2\pi C_{\theta})}}
          \nonumber \\
&&\exp\big(-\frac{1}{2}{\bf x}C_{h}^{-1}{\bf x}^{T}
             -\frac{1}{2}{\bf y}C_{\theta}^{-1}{\bf y}^{T}\big)
\delta\big(\theta-L_{i}(t)-\sqrt{W(t)}y_{t}\big)\,,
       \nonumber \\ 
\end{eqnarray} 
where the correlations matrices $C_{h}$ and $C_{\theta}$ are given by
\begin{equation}
\big(C_{h}\big)_{tt'}=\rho(x_{t},x_{t'})= 
\mbox{E}[x_{t}x_{t'}],
             \quad
\big(C_{\theta}\big)_{tt'}=\rho(y_{t},y_{t'})= 
\mbox{E}[y_{t}y_{t'}]\,.
\label{corr}
\end{equation}

Together with eqs. (\ref{defm})-(\ref{defl}) the equations (\ref{recr})-(\ref{recs}),(\ref{recM})-(\ref{recE}),(\ref{rhoh})-(\ref{corr}) form an exact recursive scheme in order to obtain the order parameters of the system. The practical difficulty that remains is the explicit correlations in the network at different time steps. As an illustration we calculate the first three time steps in Appendix A.

\section{Fixed-point equations}
\label{sec:fixp}
A second type of results can be obtained by requiring through the recursion relations (\ref{recM})-(\ref{recE}) that the local fields become time-independent. This means that most of the discrete noise part is neglected. We show that this procedure leads to the same fixed-point equations as those recently found from a replica symmetric thermodynamic approach \cite{BV02}.

First, for the BEG-model one can show that 
\begin{equation}
H(t)=-\sum_{i=1}^{N}\Big(h_{i}(t){\tilde\sigma}_{i}(t)
       +\theta_{i}(t){\tilde\sigma}_{i}^{2}(t)\Big)\,,
\end{equation}
where the set $\{\sigma_i(t)\}$ is the network configuration at time $t$ and the  $\{{\tilde\sigma}_{i}(t)\}$ are chosen such that
\begin{equation}
        \label{tilde}
     \epsilon_i[{\tilde\sigma}_{i}(t)|{\bsigma}(t)]=
         \min_{s\in \mathcal{S}} \epsilon_i[s|{\bsigma}(t)]                      \,\, 
\end{equation}
is a Lyapunov function for zero temperature. The proof is completely analogous to the argumentation used in \cite{P84} and \cite{HK94}. The choice of $\{{\tilde\sigma}_{i}(t)\}$ implies through the updating rule (\ref{eq:enpot}) that $\sigma_{i}(t+1)={\tilde\sigma}_{i}(t)$. For finite $N$, $H(t)$ is bounded from below implying that   $H(t+1)-H(t)=0$ after finitely many time steps. This can be realized for $\sigma_{i}(t+2)=\sigma_{i}(t)$ for all $i$ and, hence, both two-cycles and fixed-points satisfy this condition. We only study fixed-points.

Next, we start by eliminating the time dependence in the evolution equations for the local fields (\ref{rech})-(\ref{rect}). This leads to 
\begin{eqnarray}
h_{i}&=&\frac{1}{a}\xi_{i}m
 +\frac{1}{1-\chi_{h}}\mathcal{N}\Big(0,\frac{\alpha}{a^{2}}q\Big)
      +\frac{\alpha}{a}\eta_{h}\sigma_{i}    \\
\theta_{i}&=&\eta_{i}l
  +\frac{1}{1-\chi_{\theta}}
             \mathcal{N}\Big(0,\frac{\alpha}{a^{2}(1-a)^{2}}q\Big)
 +\frac{\alpha}{a(1-a)}\eta_{\theta}\sigma_{i}^{2} \,\,\,,
\end{eqnarray}
where from now on we forget about the pattern index $1$ and where we have defined
\begin{equation}
\eta_{x}=\frac{\chi_{x}}{1-\chi_{x}}, \quad x=h,\theta\,.
\end{equation}
This means that out of the discrete part of the local field distributions, i.e., $M_i(t)$ ($L_i(t)$), only the $\sigma_i(t-1)$ ($\sigma_i^2(t-1)$) term is kept besides, of course, the signal terms. These expressions consist out of two parts: A normally distributed part, ${\tilde h}_i=
\mathcal{N}\Big(\xi_{i}m /a,\alpha q/{a^{2}}(1-\chi_{h})^2\Big)$ and the analogous formula for ${\tilde\theta}_{i}$,
and some discrete noise part. Employing these expressions in the updating rule one finds
\begin{equation}
\sigma_{i}=g\Big({\tilde{h}_{i}} + \frac{\alpha}{a}\eta_{h}\sigma_{i},
{\tilde{\theta}_{i}}+ \frac{\alpha}{a(1-a)}\eta_{\theta}\sigma_{i}^{2}
\Big)\,.
\end{equation}
This is a self-consistent equation in $\sigma_i$ which, in general, admits more than one solution. This type of equation has been solved in the case of analog neural networks with continuous time dynamics \cite{SF93} and in the case of $Q$-Ising neural networks \cite{BJS99}, \cite{BJS98} using a Maxwell construction. Here we follow the same line of reasoning
for the joint probability distribution of the local fields in the ($h, \theta$)-plane (see fig.~1) leading to a unique solution
\begin{equation}
\sigma_{i}\equiv \tilde{g}\Big(\tilde{h}_{i},\tilde{\theta}_{i}\Big)
  = \mbox{sign}(\tilde{h}_{i})\Theta\Big(|\tilde{h}_{i}|+\tilde{\theta}_{i}
  +\Delta \Big)\,,
  \label{newg}
\end{equation}
where 
\begin{equation} 
\Delta=\frac{\alpha}{2a}\eta_h+\frac{\alpha}{2a(1-a)}\eta_{\theta}.
\label{Delta}
\end{equation}
Using the definition of the order parameters (see (\ref{eq:mdef}), (\ref{eq:ldef})) in the limit $N \rightarrow \infty$ one finds in the fixed point, dropping the index $i$
\begin{eqnarray}
m&=&\frac{1}{a}\left\langle\!\left\langle\xi \int Dz\int Dy \,\, \tilde{g}\big(\frac{1}{a}\xi m 
              +\frac{\sqrt{\alpha q}}{a(1-\chi_{h})}z,
\eta l+\frac{\sqrt{\alpha q}}{a(1-a)(1-\chi_{\theta})}y\big)
              \right\rangle\!\right\rangle
     \\
q&=&\left\langle\!\left\langle\int Dz\int Dy \,\,
\tilde{g}^{2}\big(\frac{1}{a}\xi m
+\frac{\sqrt{\alpha q}}{a(1-\chi_{h})}z,
 \eta l+\frac{\sqrt{\alpha q}}{a(1-a)(1-\chi_{\theta})}y\big)
          \right\rangle\!\right\rangle
       \\
l&=&\left\langle\!\left\langle\eta\int Dz\int Dy \,\,
\tilde{g}^{2}\big(\frac{1}{a}\xi m
+\frac{\sqrt{\alpha q}}{a(1-\chi_{h})}z,
\eta l+\frac{\sqrt{\alpha q}}{a(1-a)(1-\chi_{\theta})}y\big)
          \right\rangle\!\right\rangle\,.
\end{eqnarray}
From (\ref{recr})-(\ref{recs}),  (\ref{recD})-(\ref{recE}) and (\ref{chih})-(\ref{chit}) it is clear that 
\begin{equation}
D=\frac{q}{a^{3}(1-\chi_{h})^{2}}, \quad
E=\frac{q}{a(1-a)(1-\chi_{\theta})^{2}}
\end{equation}
with
\begin{eqnarray}
\chi_{h}&=&\frac{1}{\sqrt{\alpha a^3 D}}
   \left\langle\!\left\langle\int Dz\int Dy \,\, z\,\,
\tilde{g}\big(\frac{1}{a}\xi m+\frac{\sqrt{\alpha                                             q}}{a(1-\chi_{h})}z,
 \eta l+\frac{\sqrt{\alpha q}}{a(1-a)(1-\chi_{\theta})}y\big)
 \right\rangle\!\right\rangle
\\
\chi_{\theta}&=&\frac{1}{\sqrt{a(1-a) \alpha E}}
  \left\langle\!\left\langle\int Dz\int Dy\,\, y \,\, \tilde{g}^{2}\big(\frac{1}{a}\xi m
+\frac{\sqrt{\alpha q}}{a(1-\chi_{h})}z,
\eta l+\frac{\sqrt{\alpha q}}{a(1-a)(1-\chi_{\theta})}y\big)
     \right\rangle\!\right\rangle .
     \nonumber\\
\end{eqnarray}
These equations are the same as the fixed-point equations derived from a replica-symmetric mean-field theory treatment \cite{BV02}. 

\begin{figure}[h]
\centering\includegraphics[angle=270,scale=0.4]{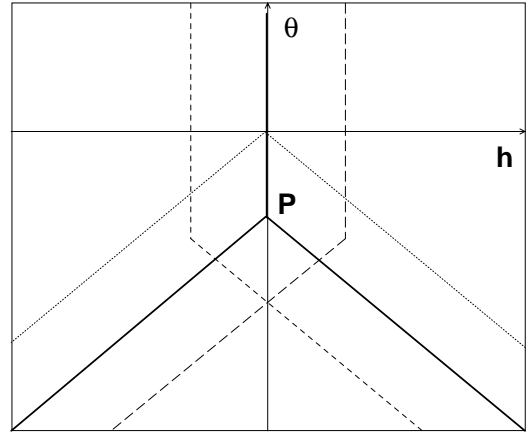}
\caption{\footnotesize The Maxwell construction. To the right of the short-dashed line the solution $\sigma=1$ exists, to the left of the long-dashed line, $\sigma=-1$ is a solution, and under the dotted line, $\sigma=0$ exists. The thick full line shows the unique solution, dividing the local fields space in three parts. The  point P is given by $P=\Big(0,-\Delta \Big)$.}
\end{figure}

\section{Local field distribution}
\label{sec:fields}
It is interesting to study the distribution of the local fields, the main ingredients in our dynamical scheme. First, we look at the stationary distribution. 
Since the Maxwell construction we used is discussed in the plane $(h,\theta)$, we want to find the  joint-distribution for the local fields, $\rho_{\infty}(h,\theta) \equiv \rho_{h(\infty),\theta(\infty)}(h,\theta)$  defined by
\begin{eqnarray}
\rho_{\infty}(h,\theta)&=& \sum_{\sigma}\int DzDy
\delta\Big(h-\frac{1}{a}\xi m-\frac{\alpha}{a}\eta_{h}\sigma-\frac{1}{1-\chi_{h}}\sqrt{\frac{\alpha q}{a^{2}}}z\Big)
\nonumber \\
&& \times \delta\Big(\theta-\eta l-\frac{\alpha}{a(1-a)}\eta_{\theta}\sigma^{2}-\frac{1}{1-\chi_{\theta}}\sqrt{\frac{\alpha q}{a^{2}(1-a)^{2}}}y\Big)        \Phi(\sigma)\,\,\,,
\end{eqnarray}
where  $\Phi(\sigma)$ is obtained from the updating rule after the Maxwell construction (see eq.~(\ref{newg}))
\begin{eqnarray}
\Phi(\sigma)&=&
\Theta\Big(-|{h}|-{\theta}-\frac{\alpha \eta_h}{2a}-\frac{\alpha \eta_{\theta}}{2a(1-a)}\Big) \delta_{\sigma,0}
 \nonumber  \\
&&+ \Theta\Big({h}\Big)\Theta\Big({h}+{\theta}
   +\frac{\alpha\eta_h}{2a}
    +\frac{\alpha \eta_{\theta}}{2a(1-a)}\Big)\delta_{\sigma,1}
   \nonumber  \\
&&+\Theta\Big(-{h}\Big)\Theta\Big(-{h}+{\theta}
 +\frac{\alpha \eta_h}{2a}
 +\frac{\alpha \eta_{\theta}}{2a(1-a)}\Big)\delta_{\sigma,-1}\,.
\end{eqnarray}

This leads to
\begin{eqnarray}
    \rho_\infty(h,\theta) &=& \rho_{+1}(h,\theta)\Theta(h-\frac{\alpha}{a}\eta_h)
                     \Theta(h+\theta-\Delta) 
         + \rho_{-1}(h,\theta)\Theta(-h-\frac{\alpha}{a}\eta_h)
                      \Theta(-h+\theta-\Delta) \nonumber \\
         &+& \rho_0(h,\theta)\Theta(-|h|-\theta-\Delta) \,,
         \label{I123}
\end{eqnarray}
where
\begin{eqnarray}
 \rho_\sigma=
\frac{(1-\chi_{h})(1-\chi_{\theta})(1-a)a^{2}}{2\pi q\alpha}
&\exp&\Big(-\frac{1}{2}\frac{(h-\frac{\xi}{a}m-\frac{\alpha}{a}\eta_{h}\sigma)^{2}}{\frac{1}{(1-\chi_{h})^{2}}\frac{\alpha q}{a^{2}}}\Big) 
\nonumber \\
&\times&\exp\Big(-\frac{1}{2}\frac{(\theta-\eta l
-\frac{\alpha}{a(1-a)}\eta_{\theta}\sigma^2)^{2}}{\frac{1}{(1-\chi_{\theta})^{2}}\frac{\alpha q}{a^{2}(1-a)^{2}}}\Big).
\end{eqnarray}

Analyzing these expressions we see that the distribution $\rho_{\infty}(h,\theta)$ shows a gap. In fig. 2 we show this gap structure which depends, of course, on the specific values of the physical parameters $\alpha, a, \chi_h, \chi_{\theta}$ of the system. The important points bordering these gaps are given by $P_{+1}=(\frac{\alpha}{a}\eta_{h},\Delta-\frac{\alpha}{a}\eta_{h})$, $P_{-1}=(-\frac{\alpha}{a}\eta_{h},\Delta-\frac{\alpha}{a}\eta_{h})$, $P_{0}=(0,-\Delta)$. 

\begin{figure}[h]
\centering\includegraphics[angle=270,scale=0.4]{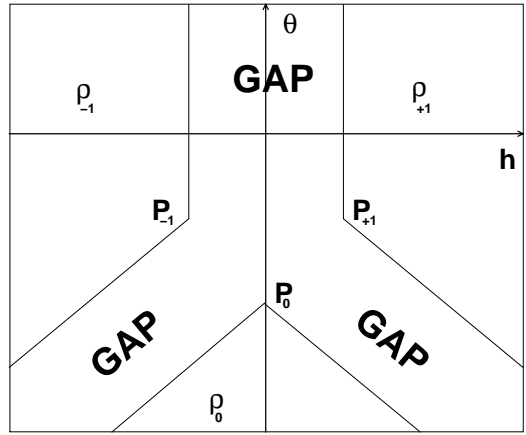}
\caption{\footnotesize 
The gap structure of $\rho_{\infty}(h,\theta)$.
The coordinates of the points $P_{+1},P_{-1},P_0$ are given in the text.
The integral  $\rho_{\pm 1}$ in (\ref{I123}) is only different from zero in the region  to the right (left) of the line on which $P_{\pm 1}$ lies;  $\rho_{0}$ exists only below the  line on which $P_{0}$ lies.} 
\end{figure} 

Dividing this joint probability by their integrations with respect to $h$ (or $\theta)$ we can obtain projections on the $\theta$ (or $h)$ axis
\begin{eqnarray}
\rho_{t=\infty}(h|\theta_{0})&=&\frac{\rho_{\infty}(h,\theta_{0})}{\int_{-\infty}^{\infty}\rho_{\infty}(h,\theta_{0})dh}
\label{projh}\\
\rho_{t=\infty}(\theta|h_{0})&=&\frac{\rho_{\infty}(h_{0},\theta)}{\int_{-\infty}^{\infty}\rho_{\infty}(h_{0},\theta)d\theta}.
\label{projt}
\end{eqnarray}

Finally, starting from  (\ref{rhoh})-(\ref{rhot}) one can write down expressions for $\rho_{h(t)}(h),\rho_{\theta(t)}(\theta)$ for the first time steps and calculate both the joint probability and its projections from it. This is illustrated in Appendix B.
All these projections will be compared with numerical simulations in the next Section.

\section{Numerical results and simulations}
\label{sec:results}
The equations derived in Sections 3-5 and Appendices A and B have been studied numerically and have been compared with simulations for systems up to $N=6000$ neurons averaged over $500$ runs.

\begin{figure}[h]
\centering\includegraphics[angle=270,scale=0.35]{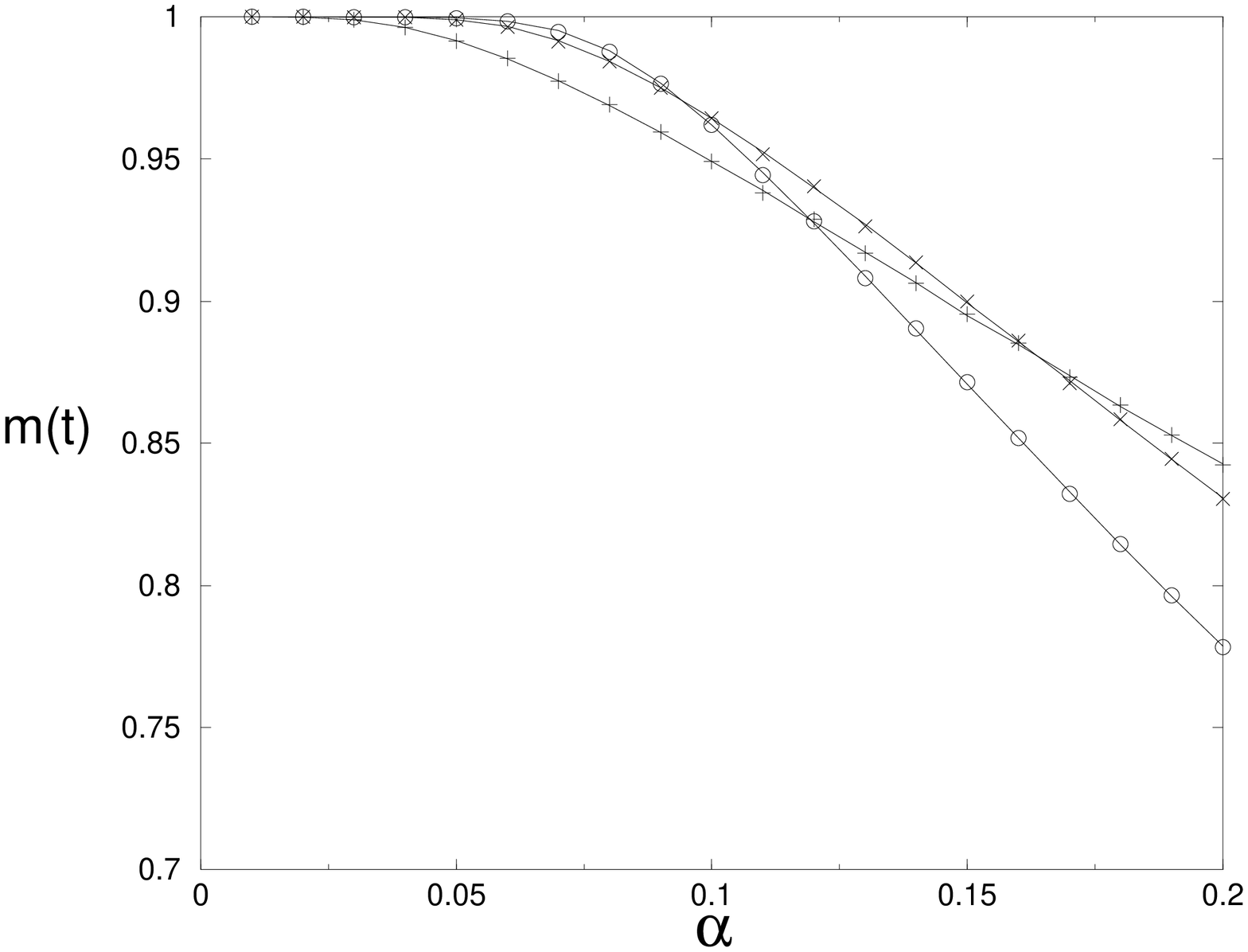}\\
\centering\includegraphics[angle=270,scale=0.35]{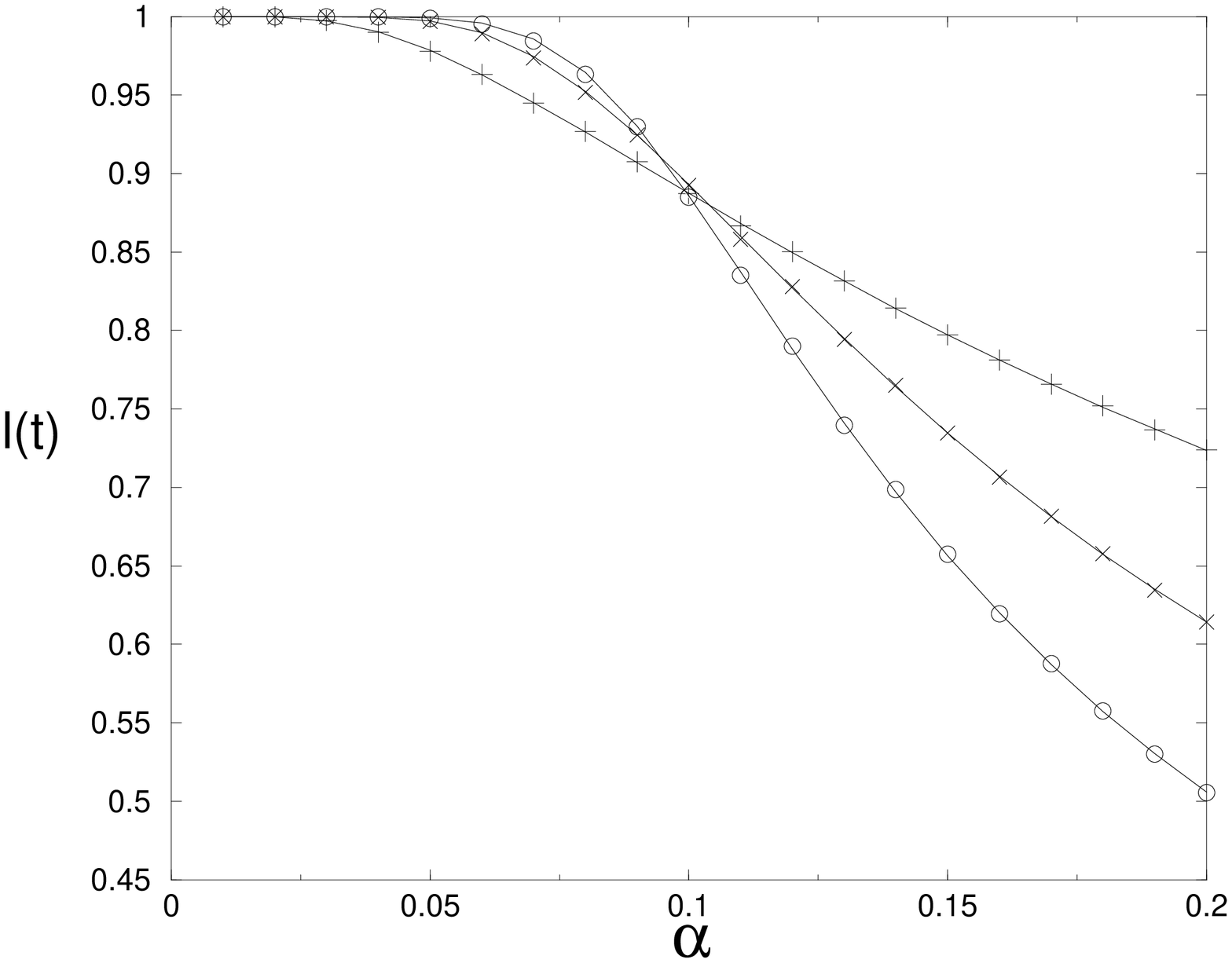}\\
\centering\includegraphics[angle=270,scale=0.35]{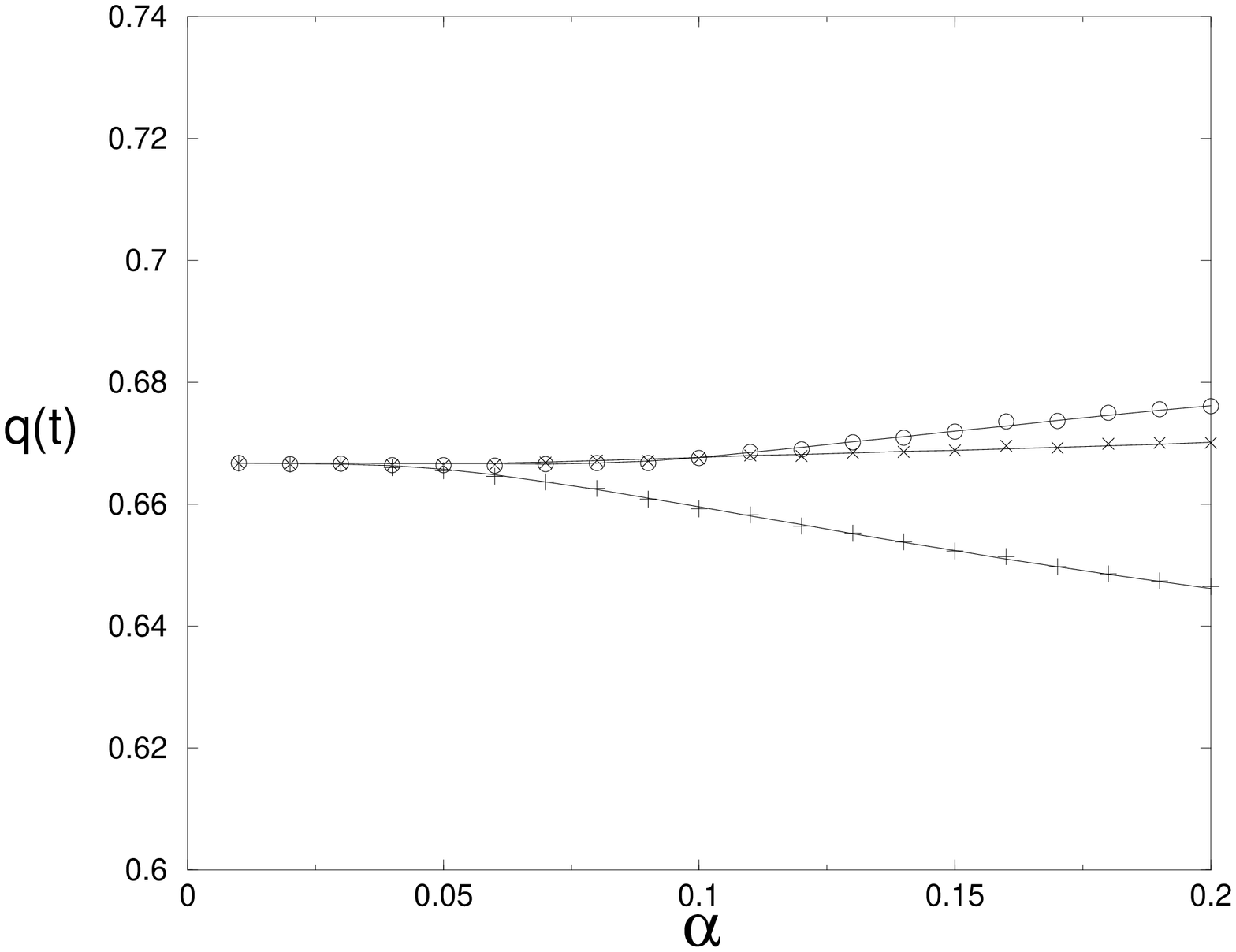}
\caption{\footnotesize Order parameters $m(t)$, $l(t)$ and $q(t)$ as a function of the capacity $\alpha$ for the first three time steps and $a=2/3$,$m_0=0.6, l_0=0.6, q_0=0.5 $. Theoretical results (solid lines) versus simulations with $N=6000$ (time 1, 2 and 3 given by the plus symbol, times symbol respectively circles) are shown. }
\end{figure}

We start with the remark that the initial conditions are not independent because positivity of the relevant probabilities implies 
\begin{eqnarray} 
 q_0>am_{0}^1, \quad \frac{m_{0}^1-q_{0}}{1-a} \ge l_0^1, \quad
 \mbox{and} &&\mbox{for} \quad a\ge q_{0}: \quad l_0^1\le                     \frac{q_{0}}{a} \nonumber \\
    && \mbox{for} \quad a\le q_{0}: \quad l_0^1 \le \frac{1-q_{0}}{1-a}\,.
    \label{cond}
\end{eqnarray}

The phase diagram of the fully connected BEG neural network has been discussed in \cite{BV02} using a replica-symmetric mean-field theory. From that work we see that for uniformly distributed patterns the critical capacity at zero temperature is $0.091$.

The first point we would like to examine is whether the recursive dynamical scheme we have derived is confirmed by simulations. This is illustrated by some typical results in fig. 3 showing the order parameters as a function of $\alpha$ for uniform patterns and $m_0=0.6, l_0=0.6, q_0=0.5 $. We see that the theoretical results, given by the explicit formula in Appendix A, and the simulations agree very well over the whole range of $\alpha$'s. Furthermore, we learn that in the retrieval regime the first time steps of the dynamics give us already a reasonable estimate for the critical capacity especially through the order parameter $l$. This is also the case for the other values of $a$, $a=0.02, 0.05, 0.08$, we have considered.

These findings are confirmed by some typical $(m(t),l(t))$ flow diagrams. In figs. 4 and 5 we show the results for uniform patterns and  two values of $\alpha$ in the retrieval region, $\alpha=0.015,0.08$, giving us a good idea of the basin of attraction. Remark that to the right of the  dotted line we cannot start initially because of the condition (\ref{cond}). At later times we can enter this region because $q(t)$ changes from its initial value $q(0)=a=2/3$. The dashed line in the figures 4a and 5a indicates the border of the basin of attraction. To have an idea about the accuracy of this basin boundary we also show in figs 4b and 5b the percentage of runs going to the attractor on the line $m_0=l_0$, starting from the (0,0)-point. The basin boundary is drawn joining the starting points of the flow lines reaching the attractor with a percentage lying between $45\%$ and $55\%$, as visualized by the two parallel dashed lines. As expected, the basin of attraction shrinks for increasing $\alpha$, due to the appearance of other thermodynamically stable states (spin-glass states),  and the error for defining the basin boundary becomes bigger. For comparable values of the relevant system parameters for the Q-Ising model one can verify (see \cite{BJS98}) that the basin of attraction of the latter is smaller. 

\vspace{0.8cm}

\begin{figure}[h]
\centering\includegraphics[angle=270,scale=0.35]{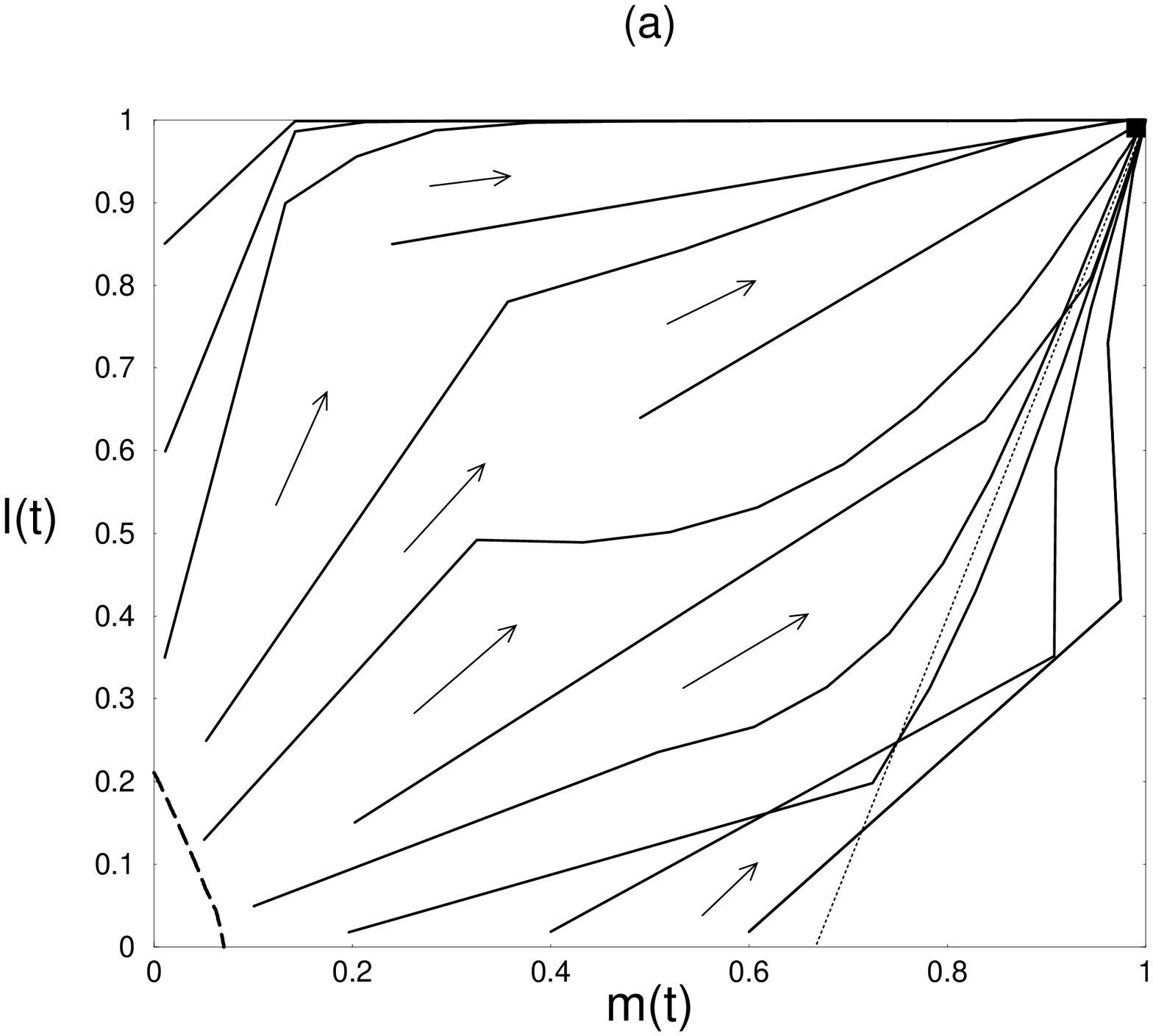}
\centering\includegraphics[angle=270,scale=0.35]{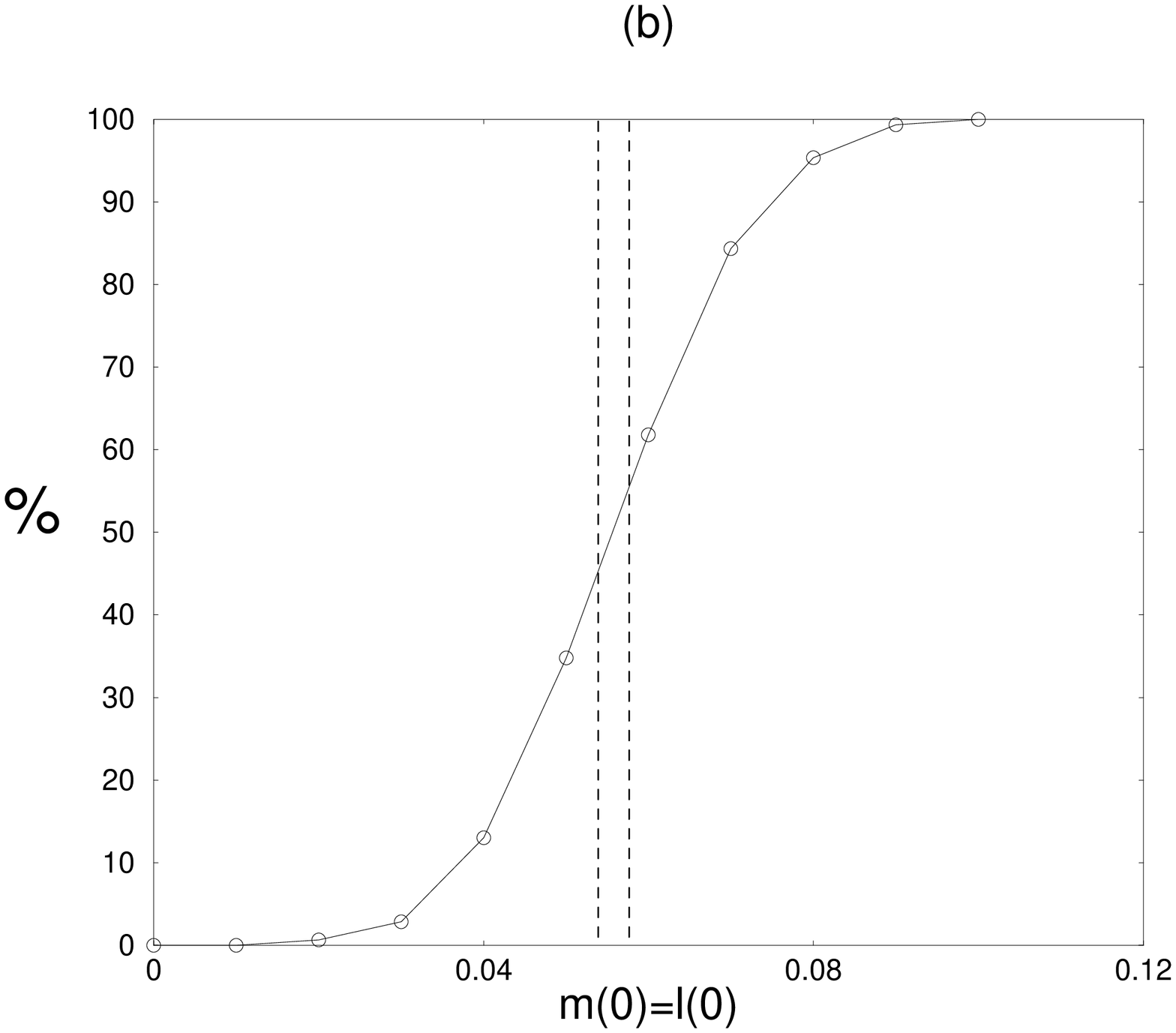}
\caption{\footnotesize Flow diagram for uniform patterns $a=2/3$ and capacity $\alpha=0.015$. In figure 4a, the long-dashed line shows the basin boundary. The parallel dashed lines in figure 4b indicate the error bounds.}
\end{figure}
\begin{figure}[h]
\centering\includegraphics[angle=270,scale=0.35]{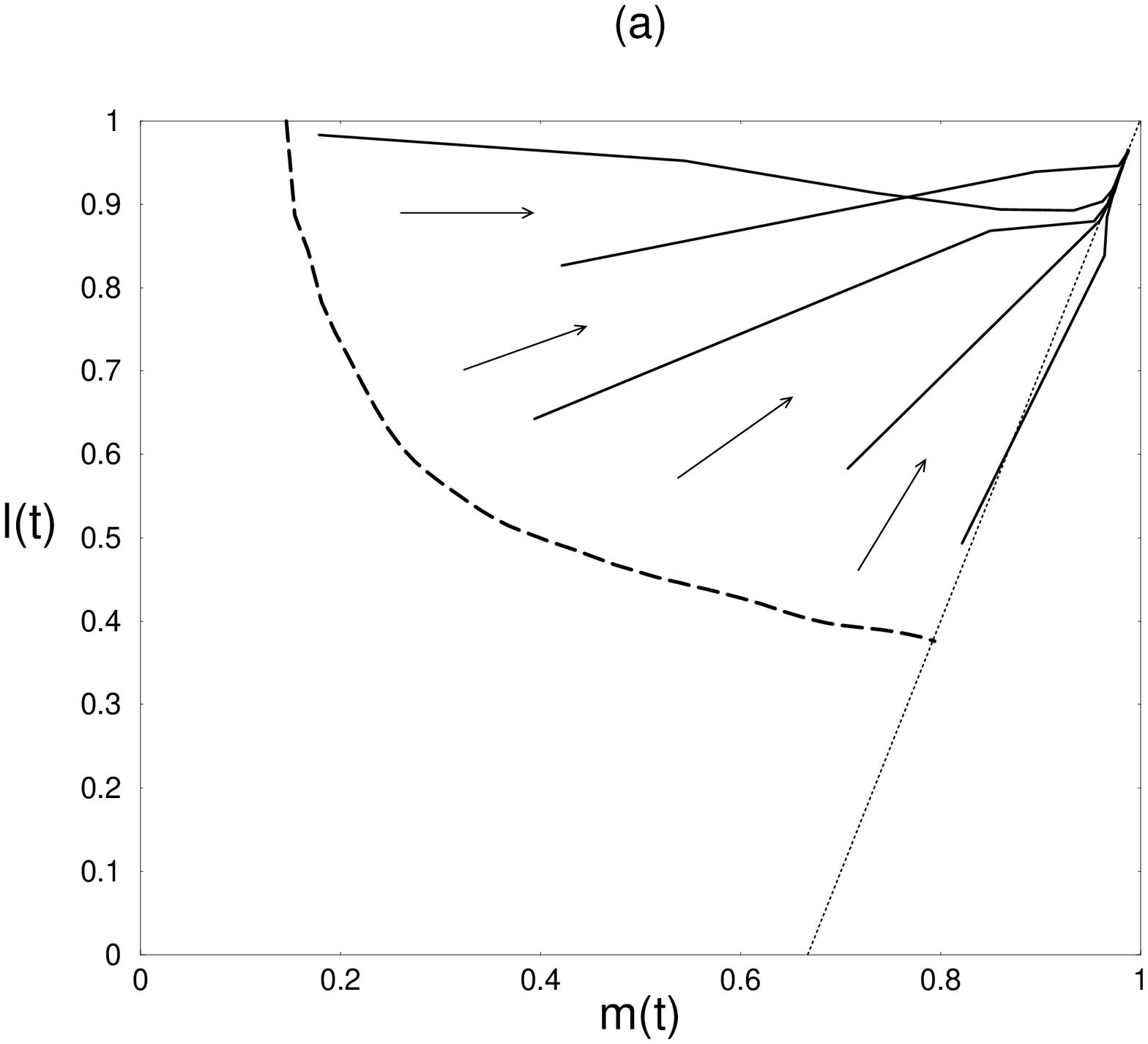}
\centering\includegraphics[angle=270,scale=0.35]{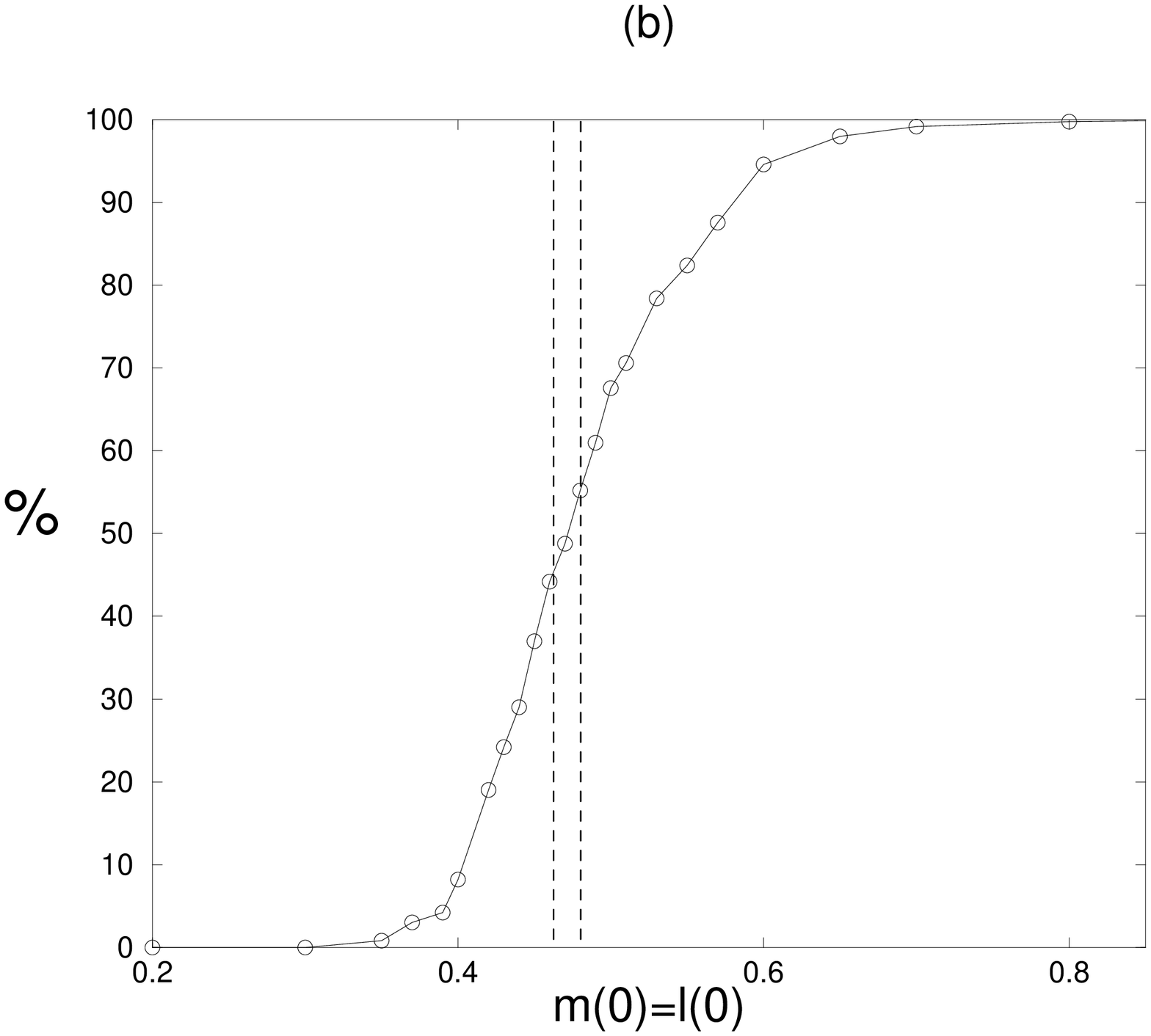}
\caption{\footnotesize  As in fig. 4 for $\alpha=0.08$. }
\end{figure}

The next point we look at is the appearance of a so-called quadrupolar state ($m=0, l \neq 0$). Extended simulations did not show any such stable state at zero temperature in the fully connected model. This is different from the findings in the diluted model where a quadrupolar state is predicted \cite{DK00}. However, recent discussions indicate that also for this diluted model the quadrupolar state is not stable at temperature zero \cite{DKTE}.
   
Finally, we consider the distributions of the local field, an important ingredient in our dynamical scheme. We have investigated them numerically using the fixed-point equations mentioned before and compared them with numerical simulations. Some typical results are shown in figs. 6 for uniform patterns. In fig. 6a we show the joint distribution of the local fields, for $\xi=0$, projected on the $h$-axis, fixing $\theta=-0.6$, in the retrieval region, $\alpha =0.064$, for $m_0=l_0=0.5, q_0=0.5$, while fig.6b represents this distribution in the non-retrieval region, $\alpha =0.13$, for $m_0=l_0=0.2, q_0=0.5$.  The first time steps, given by the explicit formula in Appendix B, are in complete agreement with the numerical simulations.

Concerning the gap structure we see that for the retrieval state there are, typically, small gaps in the equilibrium distribution. For small $\alpha$ the gaps are very narrow (see insets of fig. 6a). In the simulations these gaps show up very quickly. In order not to overload the figures we have plotted one intermediate result for the time step $t=100$. For the non-retrieval state the gaps are typically much bigger. Again in simulations the gaps show up rather quickly. In fig. 7 we plot a 3-d picture of the equilibrium distribution for $\xi=0$ in the spin-glass region, $\alpha=0.013$, for uniform patterns. The gaps are clearly visible. We recall that the theoretical equilibrium results coincide with the thermodynamic replica-symmetric solution and that it is expected that the gaps are reduced to one point for the exact solution. It is extremely difficult to find points touching the axis in the simulations because of the final size effects. Analogous results have been found for the Hopfield model, the Q-Ising model \cite{ZC89}, \cite{CS94}, \cite{BS02} and, first, in the infinite range spin-glass \cite{SK79}.

\vspace{1cm}

\begin{figure}[h]
\centering\includegraphics[angle=270,scale=0.35]{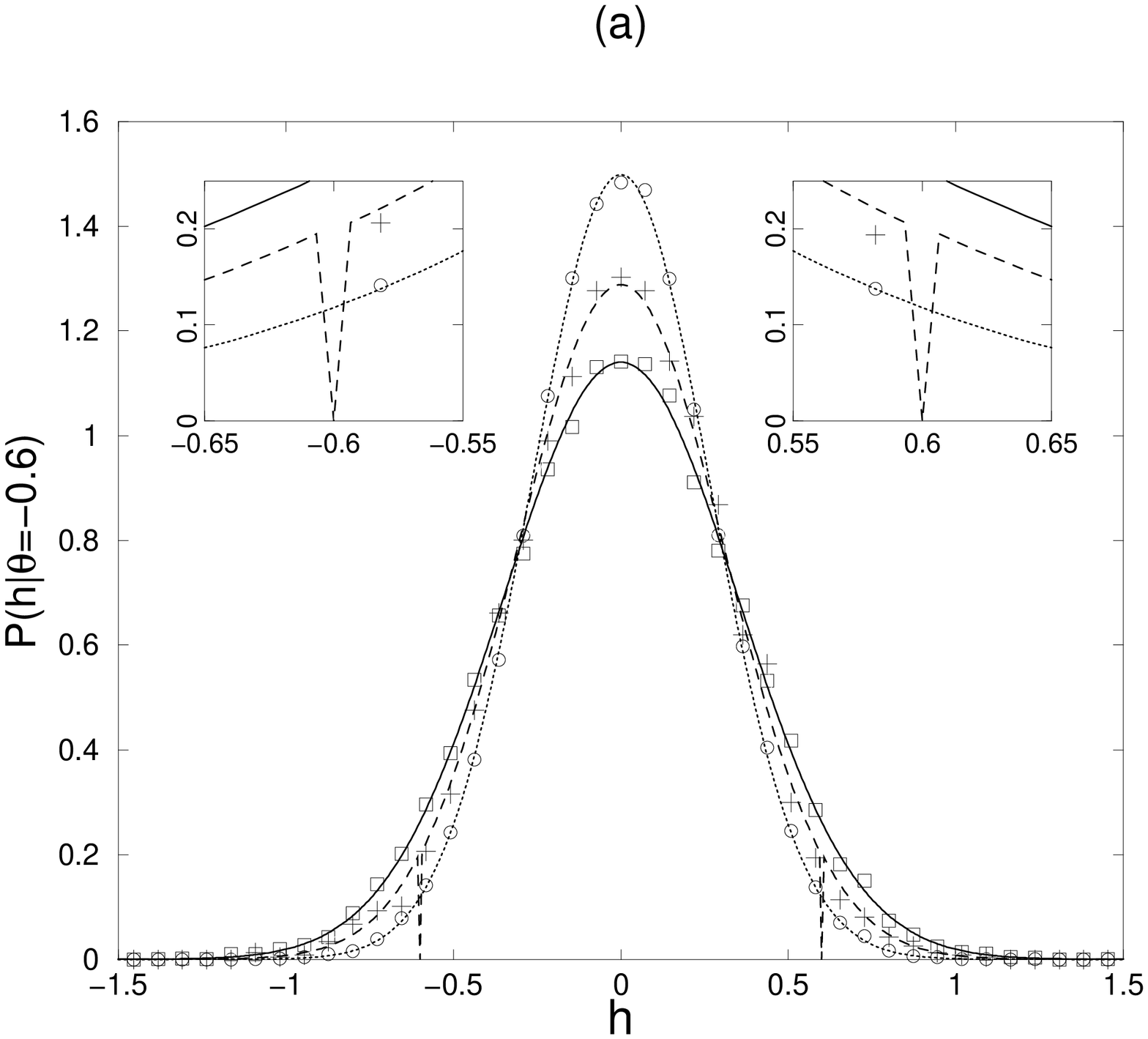}
\centering\includegraphics[angle=270,scale=0.35]{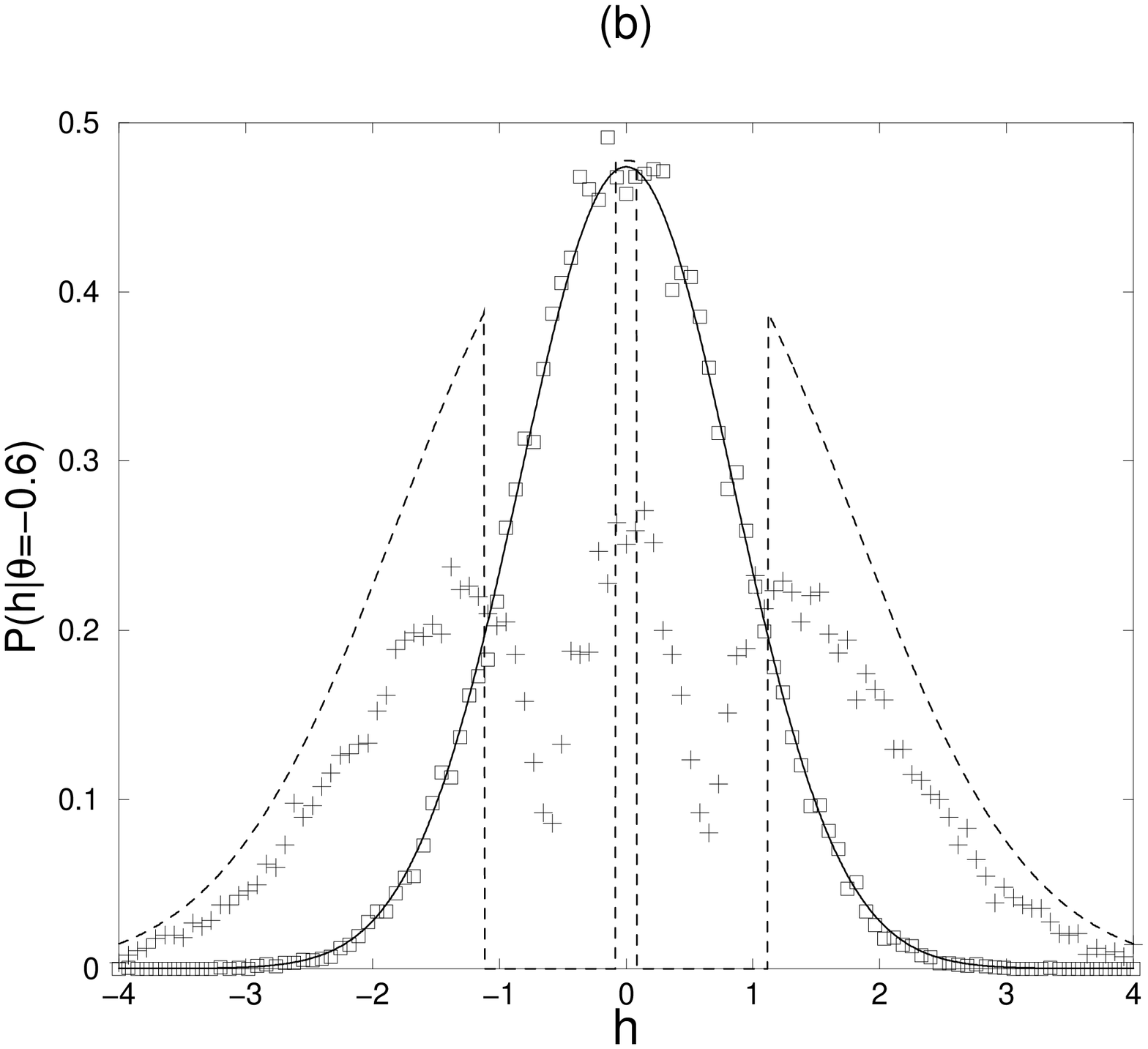}
\caption{\footnotesize Projections of the joint probability distribution of the local fields in the retrieval ($\alpha=0.064$, fig. 6a) and the spin-glass phase ($\alpha=0.13$, fig. 6b). Simulations for time steps $0$ (circles), $1$ (squares) and $100$ (plus symbol) are shown. For  clarity we have not included time $0$ results in fig. 6b. Dotted, solid and dashed lines show the theoretical results for times $0$, $1$ and $\infty$.}
\end{figure}
\begin{figure}[h]
\centering\includegraphics[angle=0,scale=0.25]{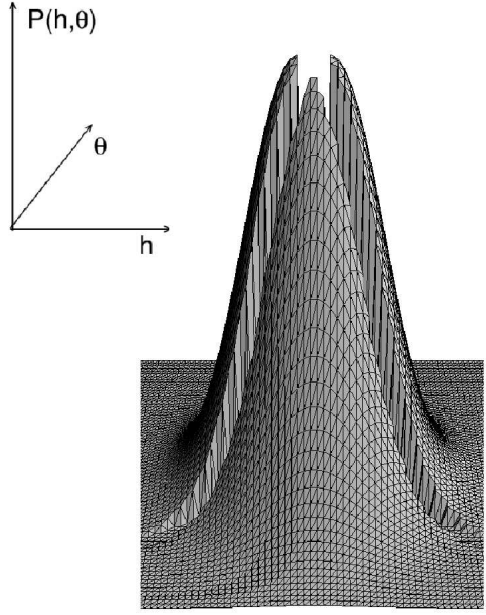}
\caption{\footnotesize A 3-d plot of the joint probability for the local field distribution for $\xi=0$ in the spin glass phase, $\alpha =0.013$, and uniform patterns, $a=2/3$, at $t\rightarrow\infty$. }
\end{figure}

\section{Conclusions}
\label{sec:con}
An evolution equation is derived for the distribution
of the local field governing the parallel dynamics at zero temperature
of BEG networks. {\it All} feedback correlations are taken into
account.  This distribution contains both a normally distributed part and  a discrete part.

Employing this evolution equation a general recursive scheme is developed
allowing one to calculate the relevant order parameters of the
system, i.e., the retrieval overlap, the activity overlap and the neural activity for {\it any} time step. This scheme has been worked out
explicitly for the first three time steps of the dynamics.

Under the condition that the local field becomes time-independent, meaning that some of the discrete noise is neglected,
fixed-point equations are obtained for the order parameters. They agree with those obtained from a mean-field replica symmetric thermodynamic approach. The gap structure of the equilibrium local field distribution is examined. The gaps in the retrieval regime are much smaller than those in the non-retrieval regime. 
  
Extensive numerical simulations are performed for a system of $6000$ neurons. They confirm the results obtained from the dynamical scheme both for the local fields and the order parameters.
Furthermore, they illustrate that the first few time steps do give a reasonable estimate of the critical capacity, especially through the activity overlap order parameter. Finally, flow diagrams indicate the size of the basin of attraction of the retrieval state as a function of the loading.

\section*{Acknowledgements}
This work has been supported in part by the Fund of Scientific Research,
Flanders-Belgium. The authors are indebted to
D.~Dominguez, R.~Erichsen~jr, I.~P\'erez Castillo, W.K.~Theumann and 
T.~Verbeiren for constructive discussions.

\section*{Appendix A}
Following the general recursive scheme developed in Section 3, evolution equations are derived for the first three time steps of the BEG fully connected network, taking into account all correlations. 
Our starting point is the set of equations for the order parameters (\ref{defm})-(\ref{defl}) with the following initial conditions
\begin{eqnarray}
&&m^{1}(0)=m_{0},\quad l^{1}(0)=l_{0},\quad q(0)=q_{0}\\
&&h_{i}(0)=\frac{1}{a}\xi_{i}^{1}m_{0}+\mathcal{N}(0,\frac{\alpha                    q_{0}}{a^{2}}),
              \quad 
\theta_{i}(0)=\eta_{i}^{1}l_{0}+\mathcal{N}(0,\frac{\alpha q_{0}}{a^{2}(1-a)^{2}})\\
&&D(0)=\frac{q_{0}}{a^{3}}, \,\,E(0)=\frac{q_{0}}{a(1-a)},\quad V(0)=\alpha aD(0),\,\, W(0)=\frac{\alpha E(0)}{a(1-a)}.
\end{eqnarray}
From now on we forget about the superscript $1$ to indicate the condensed pattern.

\subsection*{First time step}
We immediately get from (\ref{defm})-(\ref{defl})
\begin{eqnarray}
m(1)&=&\frac{1}{a}\left\langle\!\left\langle\xi\int Dz\int Dy \,\,  g\big(h'_{0}(z), \theta'_{0}(y)\big)\right\rangle\!\right\rangle
      \\
q(1)&=&\left\langle\!\left\langle\int Dz\int Dy \,\,  g^{2}\big(h'_{0}(z), \theta'_{0}(y)\big)\right\rangle\!\right\rangle
          \\
l(1)&=&\left\langle\!\left\langle\eta\int Dz\int Dy \,\, g^{2}\big(h'_{0}(z), \theta'_{0}(y)\big)\right\rangle\!\right\rangle\,,
\end{eqnarray}
where  $Dx=dx \exp({-x^{2}/2})/{\sqrt{2\pi}}$ denotes the Gaussian measure and
\begin{equation}
h'_{0}(z)=\frac{1}{a}\xi m_{0}+\sqrt{V(0)}\,z, \qquad
\theta'_{0}(y)=\eta l_{0}+\sqrt{W(0)}\,y\,\,.
\end{equation}
From the definition of $\tilde{h}_{N,i}^{\mu}(t)$ and  $\tilde{\theta}_{N,i}^{\mu}(t)$, we know that, when $N\rightarrow\infty$, the elements in the pairs $\{\xi_{i}^{\mu},g(\tilde{h}_{i}^{\mu}(0),\tilde{\theta}_{i}^{\mu}(0))\}$, $\{\xi_{i}^{\mu}, \sigma_{i}(0)\}$, $\{\eta_{i}^{\mu},g^{2}(\tilde{h}_{i}^{\mu}(0),\tilde{\theta}_{i}^{\mu}(0))\}$ and  $\{\eta_{i}^{\mu}, \sigma_{i}^{2}(0)\}$ are uncorrelated for $\mu\neq 1$. Therefore, using the recursion relations (\ref{recD}) (\ref{recE}) we get
\begin{eqnarray}
D(1)&=&\frac{q(1)}{a^{3}}+\chi_{h}^{2}(0)D(0)+2\chi_{h}(0)R(1,0)
           \\
E(1)&=&\frac{q(1)}{a(1-a)}+\chi_{\theta}^{2}(0)E(0)
                                   +2\chi_{\theta}(0)S(1,0)\,\,,
\end{eqnarray}
where, in general, the correlation parameters are defined as
\begin{eqnarray}
R(t,t')&=&\frac{1}{a^{3}}\mbox{E}\big[g(\tilde{h}(t-1),\tilde{\theta}(t-1))
              g(\tilde{h}(t'-1),\tilde{\theta}(t'-1))\big]
      \label{Rtt}    \\
R(t,0)&=&\frac{1}{a^{3}}\mbox{E}\big[\sigma(0)g(\tilde{h}(t-1),
                                           \tilde{\theta}(t-1))\big]
           \\
S(t,t')&=&\frac{1}{a(1-a)}\mbox{E}\big[g^{2}(\tilde{h}(t-1),\tilde{\theta}(t-1)                    )g^{2}(\tilde{h}(t'-1),\tilde{\theta}(t'-1))\big]
     \label{Stt}  \\
S(t,0)&=&\frac{1}{a(1-a)}\mbox{E}\big[\sigma^{2}(0)g^{2}(\tilde{h}(t-1),
                                     \tilde{\theta}(t-1))\big]
\end{eqnarray}
leading, in the limit $N \rightarrow \infty$, to the following formula for the first time step
\begin{eqnarray}
R(1,0)&=&\frac{1}{a^{3}}\left\langle\!\left\langle\sigma(0)
   \int Dz\int Dy \,\, g\big(h'_{0}(z),           \theta'_{0}(y)\big)\right\rangle\!\right\rangle
         \\
S(1,0)&=&\frac{1}{a(1-a)}\left\langle\!\left\langle\sigma^{2}(0)
    \int Dz\int Dy\,\, g^{2}\big(h'_{0}(z),       \theta'_{0}(y)\big)\right\rangle\!\right\rangle.
\end{eqnarray}
Since at time zero there are no correlations yet
\begin{eqnarray}
\chi_{h}(0)&=&\frac{1}{a\sqrt{V(0)}}
         \left\langle\!\left\langle\int Dz\int Dy \,\, z \,\,                   g\big(h'_{0}(z),                                \theta'_{0}(y)\big)\right\rangle\!\right\rangle
         \\
\chi_{\theta}(0)&=&\frac{1}{a(1-a)\sqrt{W(0)}}
        \left\langle\!\left\langle\int Dz\int Dy \,\, y \,\,                     g^{2}\big(h'_{0}(z),                          \theta'_{0}(y)\big)\right\rangle\!\right\rangle. 
\end{eqnarray}

\subsection*{Second time step}
First, we need the distribution of the local fiels at time $t=1$. This follows immediately from (\ref{recM})-(\ref{recL}) and (\ref{VWini})
\begin{eqnarray}
h_{i}(1)&=&\frac{\xi_{i}}{a}m(1)
   +\frac{\alpha}{a}\chi_{h}(0)\sigma_{i}(0)+\mathcal{N}(0,V(1))
   \label{hi1}  \\
\theta_{i}(1)&=&\eta_{i}l(1)
        +\frac{\alpha}{a(1-a)}\chi_{\theta}(0)\sigma_{i}^{2}(0)
               +\mathcal{N}(0,W(1))
               \label{ti1}\,\,.
\end{eqnarray}
These results allow us to write down the order parameters at time step $2$:
\begin{eqnarray}
m(2)&=&\frac{1}{a}\left\langle\!\left\langle \xi
\int Dz\int Dy \,\, g\big(h'_1(z), 
   \theta'_1(y) \big)\right\rangle\!\right\rangle
          \\
q(2)&=&\left\langle\!\left\langle \int Dz\int Dy \,\, g^{2}\big(h'_1(z),
\theta'_1(y)\big)\right\rangle\!\right\rangle
\\
l(2)&=&\left\langle\!\left\langle \eta\int Dz\int Dy\,\, g^{2}\big(h'_1(z),
\theta'_1(y)\big)\right\rangle\!\right\rangle\,,
\end{eqnarray}
where
\begin{eqnarray}
h'_1(z)&=&\frac{1}{a}\xi m(1)
        +\frac{\alpha}{a}\chi_{h}(0)\sigma_{i}(0)+\sqrt{V(1)}\,z \\
\theta'_1(y)&=&\eta l(1)+\frac{\alpha}{a(1-a)}\chi_{\theta}(0)\sigma_{i}^{2}(0)
         +\sqrt{W(1)}\,y
\end{eqnarray}          
        
and
\begin{eqnarray}
\chi_{h}(1)&=&\frac{1}{a\sqrt{V(1)}}
      \left\langle\!\left\langle\int  Dz\int Dy \,\, z \,\,               g\big(h'_1(z), 
        \theta'_1(y)\big)\right\rangle\!\right\rangle
    \\
\chi_{\theta}(1)&=&\frac{1}{a(1-a)\sqrt{W(1)}}\left\langle\!\left\langle
     \int Dz\int Dy \,\, y \,\, g^{2}\big(h'_1(z),
        \theta'_1(y)\big)\right\rangle\!\right\rangle .
\end{eqnarray}
The calculation of the variance of the residual overlap needs some more work. From the recursion relations (\ref{recD})-(\ref{recE}) one finds
\begin{eqnarray}
D(2)&=&\frac{q(2)}{a^{3}}+\chi_{h}^{2}(1)D(1)+2\chi_{h}(1)\big(R(2,1)
             +\chi_{h}(0)R(2,0)\big)
      \\
E(2)&=&\frac{q(2)}{a(1-a)}+\chi_{\theta}^{2}(1)E(1)
        +2\chi_{\theta}(1)\big(S(2,1)+\chi_{\theta}(0)S(2,0)\big)\,\,,
\end{eqnarray}
where $R(2,0)$ and $S(2,0)$ can be written down immediately
\begin{eqnarray}
R(2,0)&=&\frac{1}{a^{3}}\left\langle\!\left\langle\sigma(0)
 \int Dz\int Dy \,\, g\big(h'_1(z),
\theta'_1(y)\big)\right\rangle\!\right\rangle
\\
S(2,0)&=&\frac{1}{a(1-a)}\left\langle\!\left\langle\sigma^{2}(0)
\int Dz\int Dy \,\, g^{2}\big(h'_1(z),
\theta'_1(y)\big)\right\rangle\!\right\rangle .
\end{eqnarray}
To obtain $R(2,1)$ and $S(2,1)$ we remark that the local fields at time steps $0$ and $1$ are correlated. The correlation coefficients of their normally distributed part, viz.
\begin{eqnarray}
\rho_{h}(t,t')
  &=&\frac{\mbox{E}[(h(t)-M(t))(h(t')-M(t'))]}{\sqrt{V(t)}\sqrt{V(t')}}
  \\
\rho_{\theta}(t,t')
&=&\frac{\mbox{E}[(\theta(t)-L(t))(h\theta(t')-L(t'))]}
               {\sqrt{W(t)}\sqrt{W(t')}}
\end{eqnarray} 
is found using the recursion formula (\ref{hi1})-(\ref{ti1})
\begin{equation}
\rho_{h}(1,0)=\frac{R(1,0)+D(0)\chi_{h}(0)}{\sqrt{D(0)D(1)}}
\quad
\rho_{\theta}(1,0)=\frac{S(1,0)+E(0)\chi_{\theta}(0)}{\sqrt{E(0)E(1)}}\,\,.
\end{equation}
Employing all this in eqs. (\ref{Rtt}) and (\ref{Stt}) we arrive at
\begin{eqnarray}
R(2,1)&=&\frac{1}{a^{3}}\left\langle\!\left\langle
    \int D\omega_{h}^{1,0}(z,s)\int D\omega_{\theta}^{1,0}(y,t) \,\,       g\big(h'_1(z),\theta'_1(y)\big) \,\,                                           g\big(h'_{0}(s),\theta'_{0}(t)\big)\right\rangle\!\right\rangle
           \\
S(2,1)&=&\frac{1}{a(1-a)}\left\langle\!\left\langle
      \int D\omega_{h}^{1,0}(z,s)\int D\omega_{\theta}^{1,0}(y,t) \,\,               g^{2}\big(h'_1(z),\theta'_1(y)\big)\,\,                                      g^{2}\big(h'_{0}(s),\theta'_{0}(t)\big)
      \right\rangle\!\right\rangle ,
\end{eqnarray}
where the joint distribution $D\omega_{x}^{a,b}(z,y)$ equals
\begin{equation}
D\omega_{x}^{a,b}(z,y)=\frac{dzdy}{2\pi\sqrt{1-\rho_{x}^{2}(a,b)}}
\exp\left({-\frac{z^{2}-2zy\rho_{x}(a,b)+y^{2}}{2(1-\rho_{x}^{2}(a,b))}}          \right) \label{joint}\,\,.
\end{equation}

\subsection*{Third time step}
We start by writing down the distribution of the local fiels at $t=2$
\begin{eqnarray}
h_{i}(2)&=&\frac{\xi_{i}}{a}m(2) +\frac{\alpha}{a}\chi_{h}(2)
 \Big[\sigma_{i}(1)+\chi_{h}(0)\sigma_{i}(0)\Big]+\mathcal{N}(0,V(2))
       \\
\theta_{i}(2)&=&\eta_{i}l(2)+\frac{\alpha}{a(1-a)}\chi_{\theta}(1)
\Big[\sigma_{i}^{2}(1)+\chi_{\theta}(0)\sigma_{i}^{2}\Big]
          +\mathcal{N}(0,W(2))\,\,.
\end{eqnarray}
In order to write down the expressions for the order parameters starting from (\ref{defm})-(\ref{defl}) the average has to be taken over the Gaussian noise, $\sigma_i(0)$ and $\sigma_i(1)$. The average over $\sigma_i(0)$ causes no difficulties because this initial configuration is chosen randomly. The average over the Gaussian random noise variable appearing in $h_i(2)$, $\theta_i(2)$, and $\sigma_i(1)$ is more tricky because, e.g.,   $h_i(2)$ and $\sigma_i(1)$ are correlated by the dynamics. However, the evolution equation tells us that  $\sigma_i(1)$ can be replaced by $g(h_i(0),\theta_i(0))$ and, hence, its average taken over $h_i(0)$,$\theta_i(0)$ instead of $\sigma_i(1)$. From the recursion relation (\ref{rech})-(\ref{rect}) one finds for the relevant correlation coefficients 
\begin{eqnarray}
\rho_{h}(2,0)&=&\frac{R(2,0)+R(1,0)\chi_{h}(1)+D(0)\chi_{h}(1)\chi_{h}(0)}{\sqrt{D(0)D(2)}}
\\
\rho_{\theta}(2,0)&=&\frac{S(2,0)+S(1,0)\chi_{\theta}(1)+E(0)\chi_{\theta}(1)\chi_{\theta}(0)}{\sqrt{E(0)E(2)}}\,\,.
\end{eqnarray}
Using this we get
\begin{eqnarray}
m(3)&=&\frac{1}{a}\left\langle\!\left\langle\xi
 \int D\omega_{h}^{2,0}(z,s)\int D\omega_{\theta}^{2,0}(y,t)
    g\left(h'_2(z,s,t),
\theta'_2(y,s,t)\right)\right\rangle\!\right\rangle
\\
q(3)&=&\left\langle\!\left\langle
 \int D\omega_{h}^{2,0}(z,s)\int D\omega_{\theta}^{2,0}(y,t)
 g^{2}\left(h'_2(z,s,t),
   \theta'_2(y,s,t)\right)\right\rangle\!\right\rangle
 \\
l(3)&=&\left\langle\!\left\langle\eta
\int D\omega_{h}^{2,0}(z,s)\int D\omega_{\theta}^{2,0}(y,t) 
g^{2}\left(h'_2(z,s,t),
\theta'_2(y,s,t)\right)\right\rangle\!\right\rangle
\end{eqnarray}
with the joint distributions as defined before (see, (\ref{joint})) and
\begin{eqnarray}
h'_2(z,s,t)&=&\frac{1}{a}\xi m(2)
+\frac{\alpha}{a}\chi_{h}(1) [g(h'_{0}(s), \theta'_{0}(t))+  \chi_{h}(0)\sigma(0)] +\sqrt{V(2)}\,z \\
\theta'_2(y,s,t)&=&\eta l(2)+\frac{\alpha}{a(1-a)}\chi_{\theta}(1) [g^2(h'_{0}(s), \theta'_{0}(t))+ \chi_{\theta}(0)\sigma^2(0)] 
+\sqrt{W(2)}\,y\,\,.
\end{eqnarray}

In the same way further time steps can be calculated at the price of more complicated algebraic expressions.

\section*{Appendix B}
We calculate  explicitly the projected joint distributions for the local fields for the first time steps. Starting from (\ref{rhoh})-(\ref{rhot}) we obtain
\begin{eqnarray}
\rho_{h(0)}(h)&=&\frac{1}{\sqrt{2\pi V(0)}}\exp{\Big\{\frac{-(h-\frac{\xi}{a}m_{0})^2}{2V(0)}}\Big\}
\label{hint0} \\
\rho_{\theta(0)}(\theta)&=&\frac{1}{\sqrt{2\pi W(0)}}\exp\Big\{{\frac{-(\theta-\eta l_{0})^2}{2W(0)}}\Big\}
\label{tint0}
\end{eqnarray}
and
\begin{eqnarray}
\rho_{h(1)}(h)&=&\frac{1}{\sqrt{2\pi V(1)}}\exp{\Big\{\frac{-(h-\frac{\xi}{a}m(1)-\frac{\alpha}{a}\chi_{h}(0)\sigma(0))^2}{2V(1)}}\Big\}
\label{hint1}\\
\rho_{\theta(1)}(\theta)&=&\frac{1}{\sqrt{2\pi W(1)}}\exp\Big\{{\frac{-(\theta-\eta l(1)-\frac{\alpha}{a(1-a)}\chi_{\theta}(0)\sigma^{2}(0))^2}{2W(1)}}\Big\} \label{tint1}\,\,.
\end{eqnarray}

To construct the joint probability for $t=0$ we just have to take the product of (\ref{hint0}) and (\ref{tint0}). In order to find the joint probability for $t=1$ we have to average the product of (\ref{hint1}) and (\ref{tint1})over $\sigma(0)$. This leads to
\begin{eqnarray}
 \rho_{t=1}(h,\theta)&=&\frac{1}{2 \pi \sqrt{V(1) W(1)}} \nonumber \\
&& \times \Bigr[  \frac{q_{0}}{2}\exp\Big\{{\frac{-(h-\frac{\xi}{a}m(1)-\frac{\alpha}{a}\chi_{h}(0))^2}{2V(1)}}-\frac{-(\theta-\eta l(1)-\frac{\alpha}{a(1-a)}\chi_{\theta}(0))^2}{2W(1)}\Big\} \nonumber \\ && + \frac{q_{0}}{2}\exp\Big\{{\frac{-(h-\frac{\xi}{a}m(1)+\frac{\alpha}{a}\chi_{h}(0))^2}{2V(1)}}-\frac{-(\theta-\eta l(1)-\frac{\alpha}{a(1-a)}\chi_{\theta}(0))^2}{2W(1)}\Big\} \nonumber \\
&& + (1-q_{0})\exp\Big\{{\frac{-(h-\frac{\xi}{a}m(1))^2}{2V(1)}}-\frac{-(\theta-\eta l(1))^2}{2W(1)}\Big\}\Bigr]\,.
\end{eqnarray}

The projected distributions are then obtained analogously to eqs. (\ref{projh}) and (\ref{projt}). For $t=0$ we find back (\ref{hint0})
and (\ref{tint0}) confirming  again that the local fields at time zero are independent because no correlations are present yet. 
For the first time step we obtain
\begin{eqnarray}
\rho_{t=1}(h|\theta_{0})&=&
\frac{1}{\sqrt{2\pi V(1)}}
\Bigr[ \frac{q_{0}}{2}\exp\Big\{{\frac{-(\theta_{0}-\eta l(1)-\frac{\alpha}{a(1-a)}\chi_{\theta}(0))^2}{2W(1)}}\Big\} 
\nonumber \\
&&\hspace{-1cm}\times\Big(\exp\Big\{{\frac{-(h-\frac{\xi}{a}m(1)-\frac{\alpha}{a}\chi_{h}(0))^2}{2V(1)}}\Big\}+\exp\Big\{{\frac{-(h-\frac{\xi}{a}m(1)+\frac{\alpha}{a}\chi_{h}(0))^2}{2V(1)}}\Big\}\Big) \times
 \nonumber \\
&& \hspace{-1cm}+ (1-q_{0})\exp\Big\{{\frac{-(h-\frac{\xi}{a}m(1))^2}{2V(1)}}\Big\}
\exp\Big\{{\frac{-(\theta_{0}-\eta l(1))^2}{2W(1)}}\Big\}\Bigr]                  \nonumber \\
&&\hspace{-1cm}\times \Bigr[q_{0}\exp\Big\{{\frac{-(\theta_{0}-\eta l(1)-\frac{\alpha}{a(1-a)}\chi_{\theta}(0))^2}{2W(1)}}\Big\}+(1-q_{0})\exp\Big\{{\frac{-(\theta_{0}-\eta l(1))^2}{2W(1)}}\Big\}\Bigr]^{-1}\nonumber \\
\end{eqnarray}
and finally
\begin{eqnarray}
\rho_{t=1}(\theta|h_{0})&=&\frac{1}{\sqrt{2\pi W(1)}}
\Bigr[\frac{q_{0}}{2}\exp\Big\{{\frac{-(\theta-\eta l(1)-\frac{\alpha}{a(1-a)}\chi_{\theta}(0))^2}{2W(1)}}\Big\}
 \nonumber \\ 
&& \hspace{-1cm} \times\Big(\exp\Big\{{\frac{-(h_{0}-\frac{\xi}{a}m(1)-\frac{\alpha}{a}\chi_{h}(0))^2}{2V(1)}}\Big\}+\exp\Big\{{\frac{-(h_{0}-\frac{\xi}{a}m(1)+\frac{\alpha}{a}\chi_{h}(0))^2}{2V(1)}}\Big\}\Big) \nonumber \\
&& \hspace{-1cm}+ (1-q_{0})\exp\Big\{{\frac{-(h_{0}-\frac{\xi}{a}m(1))^2}{2V(1)}}\Big\}\exp\Big\{{\frac{-(\theta-\eta l(1))^2}{2W(1)}}\Big\}\Bigr]\nonumber\\
&&\hspace{-1cm}\times \Bigr[\frac{q_{0}}{2}\exp\Big\{{\frac{-(h_{0}-\frac{\xi}{a}m(1)-\frac{\alpha}{a}\chi_{h}(0))^2}{2V(1)}}\Big\}+\frac{q_{0}}{2}\exp\Big\{{\frac{-(h_{0}-\frac{\xi}{a}m(1)+\frac{\alpha}{a}\chi_{h}(0))^2}{2V(1)}}\Big\}
\nonumber\\
&&\hspace{-1cm}+(1-q_{0})\exp\Big\{{\frac{-(h_{0}-\frac{\xi}{a}m(1))^2}{2V(1)}}\Big\}\Bigr]^{-1}\,.
\end{eqnarray}

\end{document}